\documentclass[showpacs,preprintnumbers,amsmath,amssymb]{revtex4}
\usepackage{epsfig}
\usepackage{graphicx}
\usepackage{dcolumn}
\usepackage{bm}

\let\jnfont=\rm
\def\NPB#1,{{\jnfont Nucl.\ Phys.\ B }{\bf #1},}
\def\PLB#1,{{\jnfont Phys.\ Lett.\ B }{\bf #1},}
\def\EPJC#1,{{\jnfont Eur.\ Phys.\ Jour.\ C }{\bf #1},}
\def\PRD#1,{{\jnfont Phys.\ Rev.\ D }{\bf #1},}
\def\PRL#1,{{\jnfont Phys.\ Rev.\ Lett.\ }{\bf #1},}
\def\MPLA#1,{{\jnfont Mod.\ Phys.\ Lett.\ A }{\bf #1},}
\def\JPG#1,{{\jnfont J.\ Phys.\ G}{\bf #1},}
\def\CTP#1,{{\jnfont Commun.\ Theor.\ Phys.\ }{\bf #1},}
\def\ZPC#1,{{\jnfont Z.\ Phys.\ C }{\bf #1},}
\def\JHEP#1,{{\jnfont JHEP \ }{\bf #1},}
\def\q_slash{\not{\hbox{\kern-2.1pt $q$}}}
\def\p_slash{\not{\hbox{\kern-4.0pt $p$}}}
\def\k_slash{\not{\hbox{\kern-2.1pt $k$}}}

\begin{document}

\preprint{\parbox{1.2in}{\noindent CUMQ/HEP~144 }}

\title{\ \\[10mm]
SUSY-induced FCNC top-quark processes at the Large Hadron Collider
}

\author{ J. J. Cao$^1$, G. Eilam$^1$, M. Frank$^2$, K. Hikasa$^3$,
        G. L. Liu$^4$, I. Turan$^2$, J. M. Yang$^5$ \\~~
}
\affiliation{
$^1$ Physics Department, Technion, 32000 Haifa, Israel\\
$^2$ Department of Physics, Concordia University, Montreal, H4B 1R6, Canada\\
$^3$ Department of Physics, Tohoku University, Sendai 980-8578, Japan\\
$^4$ Service de Physique Theorique CP225, Universite Libre de
          Bruxelles, 1050 Brussels, Belgium \\
$^5$ Institute of Theoretical Physics, Academia Sinica, Beijing 100080, China
}

\begin{abstract}
We systematically calculate various flavor-changing neutral-current top-quark processes
induced by supersymmetry at the Large Hadron Collider, which include five decay modes and six
production channels. To reveal the characteristics  of these
processes, we first  compare the dependence of
the rates for these channels on the relevant supersymmetric parameters,
then we scan the whole parameter space to find their maximal
rates, including all the direct and indirect current
experimental constraints on the scharm-stop flavor mixings. We
find that, under all these constraints, only a few channels,
 through $c g \to t $ at parton-level and $t \to c
h $, may be observable at the Large Hadron Collider.
\end{abstract}

\maketitle

\section{\bf Introduction}

The study of various flavor-changing neutral-current (FCNC)
processes has been  shown to be  very useful.  In particular, as
the heaviest fermion in the standard model (SM), the top quark may play a special
role in FCNC phenomenology. In  the SM the FCNC interactions of
the top quark are extremely suppressed by the GIM mechanism, so
that no FCNC top quark rates can reach an observable level at
current or future colliders \cite{tcvh-sm,tcgg-sm,eetc-sm}. Thus,
the observation of any FCNC top quark process would be a robust
evidence for new physics beyond the SM.  Due  to its heaviness,
the top quark is very sensitive to new physics. Indeed, several models beyond the
SM often predict much larger FCNC top quark
interactions \cite{Larios:2006pb}. Such FCNC interactions can induce various
top quark production and decay channels, which can be explored in future
collider experiments
\cite{Aguilar-Saavedra:2004wm,Aguilar-saavedra:linear} and serve
as a good probe for new physics.

So far, much effort has been spent on the exploration of the FCNC
top quark interactions. On the experimental side, the Tevatron CDF
and D0 collaborations have reported interesting bounds on the FCNC
top quark decays from Run 1 experiment and will tighten the bounds
from the ongoing Run 2 experiments \cite{cdfd0}. On the
theoretical side, various FCNC top quark decays and top-charm
associated productions at high energy colliders were extensively
studied in the SM \cite{tcvh-sm,tcgg-sm,eetc-sm}, the Minimal
Supersymmetric Standard Model (MSSM)
\cite{tcv-mssm,tch-mssm,eetc-mssm,pptc-mssm,Gad-pptc-mssm} and
other new physics models \cite{tc-TC2,tcv-TC2,other}. These
studies showed that the SM predictions for such
processes are far below the detectable level. However, some new
physics can enhance them by several orders of magnitude, which makes
them potentially accessible at future colliders.

The Large Hadron Collider (LHC) at CERN will be a powerful
machine for studying the top quark properties such as its FCNC
interactions since it will produce top quarks copiously. Analysis
\cite{Aguilar-Saavedra:2004wm} showed that some FCNC top quark
rare decays with branching ratios as low as ${\cal O}(10^{-5})$
could be accessible at the LHC. Due to its high energy and high
luminosity, the LHC will be the main utility for exploring
FCNC top quark production channels \cite{Aguilar-Saavedra:2004wm}.

The minimal supersymmetric standard model (MSSM)  is a leading candidate for new
physics beyond the SM and its consequences will be extensively explored at the LHC.
In this paper, we throughly
investigate various top quark FCNC processes in the
framework of the MSSM.  A characteristic feature of the model is that, in addition to the
FCNC interactions generated at loop level by the CKM mixing matrix
in the SM, it predicts FCNC interactions from soft SUSY breaking
terms \cite{susyflavor}. These additional FCNC interactions depend
on the squark flavor mixings, and for the case of top quark, they
are sensitive to the potentially large mixings between charm
squarks (scharms) and top squarks (stops).

In this paper, we will examine six production channels, which proceed through the parton-level:
\begin{eqnarray}
&& g g \to t \bar{c}, \label{pro-1} \\
&& c g \to t, \\
&& c g \to t g,  \\
&& c g \to t Z, \\
&& c g \to t \gamma,\\
&& c g \to t h,
\label{processes}
\end{eqnarray}
and five FCNC top quark decay modes:
\begin{eqnarray}
t & \to & c g,  \label{decy-1}\\
t & \to & c g g,\\
t & \to & c Z, \\
t & \to & c \gamma,\\
t & \to & c h,
\label{fcncdecays}
\end{eqnarray}

Among these, the decays $t \to c g, cgg, c Z, c \gamma, c h$
and the production $ g g \to t \bar{c}$ have already been studied
in the MSSM  before \cite{tcv-mssm,pptc-mssm,Gad-pptc-mssm}, while the others
have not been studied so far. Here we
perform a comprehensive study of all these processes in the MSSM
for the following purposes:
\begin{itemize}
\item[(1)] Since in the MSSM all these processes are induced
mainly by scharm-stop mixings and each of them involves the same
set of SUSY parameters, they are correlated. Although
some of them have been studied in the literature, they were treated
individually in different papers. Even though a combined analysis has been
done in \cite{Aguilar-Saavedra:2004wm} within the framework of effective Lagrangian where the
coefficients of all FCNC interactions are independent, such analysis is
missing in an explicit model.
Therefore, a comprehensive and comparative study of all these processes in an explicit model is necessary.
\item[(2)] Through a comparative study of all these
channels, we could determine the relative size of their rates. This is
useful since the LHC experiments can in principle measure
each of them and the pattern of relative rates can be tested. Only
by considering all these processes  together, we might know which
one has the largest rate and will hopefully  be discovered at
the LHC, if the MSSM is the correct framework.
\item[(3)] By
scrutinizing the dependence of the rates of these transitions on the
relevant SUSY parameters, one can determine the most sensitive parameters
of the model and then discuss how the future LHC measurements could
possibly bound them.

\item[(4)] Performing the scan over the whole parameter
space, subject to all the direct and indirect current experimental
constraints on the scharm-stop flavor mixings \cite{cao}, the maximal
rate for each process can be determined and, in this way, we can
pinpoint which ones are hopefully observable at the LHC. Of
course, this does not mean that one should give up searching
for those low-rate processes at the LHC. As stated above, the LHC
measurements can readily place bounds on the sensitive SUSY
parameters and such bounds are complementary to the current
experimental constraints, most of which are indirect constraints.
\end{itemize}
This paper is organized as follows. In Sec. II we discuss the
possible sources of flavor violation in the MSSM and give the FCNC
interaction Lagrangian relevant to our calculations. In Sec. III
we introduce a method to calculate various top quark FCNC
processes. This method, as will be shown, can greatly simplify our
calculations. The predictions of the rates are given in Sec. IV,
with emphasis on  illustrating the
characteristics  of their dependence on the relevant SUSY
parameters. In Sec. V we consider various experimental constraints
on the sources of flavor violation and scan the parameter space to
find the maximal rates at the LHC.  We  draw our conclusion in
Sec. VI.  Finally, we give the expressions for the loop results
in the Appendix.

\section{\bf FCNC interactions in the MSSM}

There are two sources of flavor
violation in the MSSM \cite{susyflavor}. The first one arises from the
flavor mixings of up-quarks and down-quarks, which are
described by the CKM matrix (inherited from the SM).
The second one results from the misalignment between the rotations
that diagonalize the quark and squark sectors due to the presence of
soft SUSY breaking terms. This source can induce large top quark
FCNC processes and is the focus of investigation in this paper.

In the super-CKM basis with states ($\tilde u_L$, $\tilde c_L$,
$\tilde t_L$, $\tilde u_R$, $\tilde c_R$, $\tilde t_R$) for
up-squarks and ($\tilde d_L$, $\tilde s_L$, $\tilde b_L$, $\tilde
d_R$, $\tilde s_R$, $\tilde b_R$) for down-squarks, the $6\times
6$ squark mass matrix ${\cal M}^2_{\tilde q}$ ($\tilde q=\tilde u,
\tilde d$) takes the form \cite{susyflavor}
\begin{eqnarray}
{\cal M}^2_{\tilde q}=\left( \begin{array}{ll}
  (M^2_{\tilde q})_{LL}+ m_q^2 + \cos 2\beta M_Z^2 ( T_3^q - Q_q
s_W^2) \hat{\mbox{\large 1}} &\;\;\;(M^2_{\tilde q})_{LR}- m_q \mu (\tan \beta)^{- 2 T_3^q}\\
  (M^2_{\tilde{q}})_{LR}^\dag - m_q \mu (\tan \beta)^{- 2 T_3^q}
             &\;\;\;(M^2_{\tilde{q}})_{RR}+ m_q^2 + \cos
2\beta M_Z^2 Q_q s_W^2 \hat{\mbox{\large 1}}  \end{array} \right)
, \label{sq-matrix}
\end{eqnarray}
where the soft mass parameters $(M^2_{\tilde q})_{LL}$, $(M^2_{\tilde
q})_{LR}$ and $(M^2_{\tilde q})_{RR}$ are $3 \times 3$ matrices in
flavor space, $\hat{\mbox{\large 1}}$ stands for the unit matrix,
$m_q$ is the diagonal quark mass matrix, $T_3^q=1/2$ for
up-squarks and $T_3^q=-1/2$ for down-squarks, and $\tan \beta
=v_2/v_1$ is the ratio of the vacuum expectation values of the
Higgs fields. In general, the soft mass parameters are flavor
non-diagonal. Since the low energy experimental data, such as
$K^0-\bar K^0$, $D^0-\bar D^0$ and $B^0_d-\bar B^0_d$ mixings,
require the flavor mixings involving the first generation
squarks to be negligibly small \cite{susyflavor}, we only consider
the flavor mixings of the second and third generations and
parametrize the soft mass parameters as
\begin{eqnarray}
(M^2_{\tilde{u}})_{LL} &= & \left ( \begin{array}{ccc}
    M_{Q_1}^2 &  0                           &  0 \\
    0         &  M_{Q_2}^2                   & \delta_{LL} M_{Q_2} M_{Q_3} \\
    0         &  \delta_{LL} M_{Q_2} M_{Q_3} & M_{Q_3}^2 \end{array}  \right ), \nonumber \\
 (M^2_{\tilde{u}})_{LR} &=&  \left (
\begin{array}{ccc}
0     &  0   &  0 \\
0     &  0   & \delta_{LR} M_{Q_2} M_{U_3}\\
0     &  \delta_{RL} M_{U_2} M_{Q_3}  & m_t A_t \end{array}  \right ), \nonumber \\
 (M^2_{\tilde{u}})_{RR} &= & (M^2_{\tilde{u}})_{LL}|_{M_{Q_i}^2 \to M_{U_i}^2,~ \delta_{LL} \to
 \delta_{RR}}, \label{up squark}
\end{eqnarray}
for up-type squarks. Similarly, for down-squarks  we have
\begin{eqnarray}
(M^2_{\tilde{d}})_{LR}& = & \left ( \begin{array}{ccc}
0          &  0                  &  0 \\
0                                &  0   & \delta_{LR}^d M_{Q_2} M_{D_3}\\
0 &  \delta_{RL}^d M_{D_2} M_{Q_3}  & m_b A_b \end{array}  \right
),  \nonumber  \\
(M^2_{\tilde{d}})_{RR} &= & (M^2_{\tilde{u}})_{LL}|_{M_{Q_i}^2 \to
M_{D_i}^2,~ \delta_{LL} \to \delta_{RR}^d}. \label{delta'-LR}
\end{eqnarray}
Due to $SU_L(2)$ gauge invariance, $(M^2_{\tilde d })_{LL}$ is
given by
\begin{eqnarray}
(M^2_{\tilde{d}})_{LL} = V_{CKM}^\dag (M_{\tilde{u}}^2)_{LL}
V_{CKM}.  \label{SU2}
\end{eqnarray}
Note that the mixing parameters, $\delta^d$ in the down sector defined in
Eq.~(\ref{delta'-LR}) are independent of $\delta$ in the up sector defined in
Eq.~(\ref{up squark}), and in general, $\delta_{LR} \neq
\delta_{RL}$. For the diagonal elements of left-right mixings in
Eq.~(\ref{up squark}) and Eq.~(\ref{delta'-LR}), we only kept the
terms of third-family squarks, since we adopted the popular
assumption that they are proportional to the corresponding quark
masses.

It is clear that the mixing parameters in the squark mass matrices
affect both the squark mass and its interactions. For example, in
the presence of flavor mixings, squark-quark interactions are given by
\begin{eqnarray}
V(\bar{q} X \tilde{q}_{\alpha}^{\prime}) \; = \;
 \Gamma_{q}^{i \alpha} \;
 V(\bar{q} X \tilde{q}^{\prime}_i)~,  \label{interaction}
\end{eqnarray}
where $V(\bar{q} X \tilde{q}_{\alpha}^{\prime})$ denotes the
interaction in squark mass-eigenstates, $V(\bar{q} X
\tilde{q}^{\prime}_i)$ is that in the interaction basis, $X$ may
be gluino, neutralino or chargino,
and $\Gamma_q$ is the unitary matrix which diagonalizes the squark
mass matrix. For the convenience in the following discussions,
we give the interaction Lagrangian for up-type quarks
\cite{susyflavor,mssmfeynmanrule}:
\begin{eqnarray}
{\cal{L}}_{u \tilde{u} \tilde g}&=& \sum_{i=1}^{3}\sqrt{2} g_s \,
T^a_{st} \left[ \bar u^{s}_i \,(\Gamma_U)^{i \alpha} P_L\, \tilde
g^a \,\tilde u^{t}_\alpha - \bar u^{s}_i \,(\Gamma_U)^{(i+3)
\alpha}\,P_R \,\tilde g^a \,\tilde u^{t}_\alpha + \text{h.c.} \right]\, , \label{interaction1}  \\
{\cal{L}}_{u\tilde{u}\tilde{\chi}^{0}}&=&\sum_{n=1}^{4}\sum_{i=1}^{3}
\frac{g}{\sqrt{2}} \left\{ \bar{u}_{i}\,N_{n1}^{*}\,\frac{4}{3}
\tan \theta _{W} \,P_L\,\tilde{\chi}_{n}^{0}\,(\Gamma_U)^{(i+3)
\alpha}\,\tilde{u}_\alpha -\bar{u}_{i}\,N_{n4}^{*}\,
\frac{(m_u)_{ij}}{M_W \sin \beta}
\,P_L\,\tilde{\chi}_{n}^{0}\,(\Gamma_U)^{j \alpha
}\,\tilde{u}_\alpha
\right.   \nonumber \\
&&- \left.\bar{u}_{i}\, \left( N_{n2}+  \frac{1}{3}N_{n1}\tan
\theta _{W}\right)\,P_R \,\tilde{\chi}_{n}^{0}\,(\Gamma_U)^{i
\alpha}\,\tilde{u}_\alpha -\bar{u}_{i}\,N_{n4}\,
\frac{(m_u)_{ij}}{M_W \sin \beta}
\,P_R\,\tilde{\chi}_{n}^{0}\,(\Gamma_U)^{(j+3) \alpha}
   \tilde{u}_\alpha \right\} \,,  \\
{\cal{L}}_{u\tilde{d}\tilde{\chi}^{+}} &= &\sum_{\sigma=1}^{2}\,
\sum_{i,j=1}^{3} g \left\{ \bar{u} _ {i}\,[V_{\sigma
2}^{*}\,(\frac{m_u}{\sqrt{2} M_W \sin \beta} V_{CKM})_{ij}]
\,P_L\,\tilde{\chi} _{\sigma}^{+}\,(\Gamma_D)^{j \alpha
}\,\tilde{d}_{\alpha}-\bar{u}_{i} [ U_{\sigma 1} (V_{CKM})_{ij}]
P_R\,
\tilde{\chi}_{\sigma}^{+}\,(\Gamma_D)^{j \alpha} \,\tilde{d}_\alpha \right.   \nonumber \\
& &  \left. +\,\bar{u}_{i}\,[U_{\sigma 2}\,(V_{CKM}
\frac{m_d}{\sqrt{2} M_W \cos \beta})_{ij}]
\,P_R\,\tilde{\chi}_{\sigma}^{+}\,(\Gamma_D)^{(j+3) \alpha
}\,\tilde{d}_\alpha \right\} +\text{h.c.} \,, \label{interactions}
\end{eqnarray}
where $T^{a}$ are the $SU(3)_{c}$ generators, $i=1,2,3$ is the
generation index, $\alpha =1, \ldots, 6$ is the squark
flavor index, $s$ and $t$ are color indices, $N$ is the $4\times 4$
rotation matrix defined by $N^{*}M_{\tilde
\chi^0}N^{-1}=\mathrm{diag}(m_{\tilde {\chi}_{1}^{0}},\,m_{\tilde
{\chi}_2^0}, \,m_{\tilde {\chi}_3^0}, \,m_{\tilde {\chi}_4^0})$,
 the index $\sigma$ refers to chargino mass eigenstates, and $V$
and $U$ are the usual chargino rotation matrices defined by
$U^{*}M_{\tilde {\chi} ^{+}}V^{-1}=\mathrm{diag} (m_{\tilde {\chi}
_{1}^{+}},m_{\tilde {\chi} _{2}^{+}})$. From the above interactions
one can see that  FCNC neutralino and
gluino interactions only arise from up-type squark
mixings, while the  FCNC chargino interactions are induced from both the off-diagonal elements in the CKM matrix, and
from the flavor mixings in down-type squark mass matrix.

Although each of the above interactions contributes to the top quark
FCNC transitions by  gaugino mediated loops, the contributions could be of quite
different magnitude.
Since both the neutralino and gluino contributions depend on the
same parameters in the up-type squark mass matrix, their different
coupling strength indicate that the neutralino contribution is
much smaller than the gluino contribution, except for a very massive gluino and light neutralino
scenario.
Noting that B-physics requires small
 $\delta^d \leq {\cal O}(0.1)$ \cite{bsr,Ball,B-summary},
the FCNC interactions induced by charginos are in general not
large. Recently, these three types of contributions to $ g g \to t
\bar{c}$ at the LHC were simultaneously calculated
in \cite{Gad-pptc-mssm}, and it was shown that both the neutralino
and the chargino contribution are several orders  of
magnitude smaller than the gluino contribution for most SUSY
parameter space. Since we are
mostly interested in the parameter regions with large predictions for FCNC processes,
in this paper we consider only the
gluino-mediated contributions.

Even if only the gluino-mediated loops are considered in calculating the
top quark FCNC interactions, the model has still a large parameter set.
Beside the gluino mass, there are nine
soft mass parameters in the scharm-stop mass matrix, which
complicates our analysis.
In order to simplify our calculations, we neglect
the charm quark mass. Then the  amplitude squared for any top
quark FCNC mode/channel considered in this paper can be decomposed as
\begin{eqnarray}
|M|^2 = |M|^2_L + |M|^2_R,   \label{decompose}
\end{eqnarray}
where $|M|^2_L$ ($|M|^2_R$) is the amplitude squared with a left-handed (right-handed)
charm quark, as either an external state or internal state in Feynman diagrams.
Furthermore, by
using the mass insertion method\cite{mass insertion}, one can easily
find that $|M|^2_L$ ($|M|^2_R)$ vanishes if there is no
left-handed (right-handed) scharm mixings with stop, and that these amplitudes have
a weaker dependence on right-handed (left-handed) scharm
mixings than on the left-handed (right-handed) scharm mixings.
These features, verified numerically by our calculations,
motivate us to consider the case with only left-handed scharm mixings in top quark
FCNC processes. In this case, the relevant soft mass
parameters are reduced to seven, since we set $\delta_{RL}\delta_{RR} =0 $ and $M_{U_2}^2 $ is irrelevant to our
calculation (see Eq.~(\ref{up squark})). Throughout this paper, we
always consider this case, but we note that for those transitions
not involving $W, Z$ bosons, the results for $|M|^2_L $ can be
applied to $|M|^2_R $ with the substitutions
$R \leftrightarrow L$ and $M_{U_i} \leftrightarrow M_{Q_i}$.

Another reason for considering only left-handed scharm mixings with stops
is that these are well motivated in popular flavor-blind
SUSY breaking scenarios, such as the mSUGRA model \cite{sugra} and
gauge-mediated SUSY-breaking models \cite{gmsb}. In these models,
the sfermion-mass matrices are flavor diagonal at the
SUSY-breaking scale, but the Yukawa couplings can induce flavor
mixings when evolving the matrices down to the electroweak scale.
Estimates of these radiatively induced off-diagonal squark-mass
terms indicate the magnitude for left-handed flavor mixings are
proportional to bottom quark mass, while those for the right-handed
scharm are proportional to charm quark mass \cite{hikasa}.
Therefore, in phenomenological studies of scharm-stop mixings, one
usually assumes the existence of left-handed scharm mixings.

Finally, it should be pointed out that although we make use of the
parametrization in Eqs.(12-14), which is widely used in the literature
for the calculations by mass insertion approximation,
our calculations are the full computation in the mass eigentsate basis
of squarks (that is we first diagonalize the squark mass matrices
and then perform the loop calculations in the mass eigentsate basis).
Such a treatment, unlike the mass insertion method \cite{mass insertion}
which makes sense only for $\delta's < 1$,  can allow for $\delta's > 1$
(as will be shown in Sec. V, in some cases $\delta's > 1$ can be permitted by
all experimental constraints because we use non-universal squark mass
parameters, that is $M_Q$, $M_U$ and $M_D$ are not degenerate).
Note that although all our numerical results are obatined from
such full computation in the mass eigenstate basis,
we will utilize the mass insertion method when we try to
qualitatively explain the behaviors of some results.

\section{\bf The Effective Vertex Method }

We  introduce a method which can  greatly simplify our
calculations since it avoids repetition of the evaluation of a same
loop-corrected vertex in different places, or in
different processes. All results in this paper were obtained by
this method, and some of them were cross-checked by other tools
such as {\tt FormCalc} \cite{Hahn}.

\begin{figure}[hbt]
\begin{center}
\epsfig{file=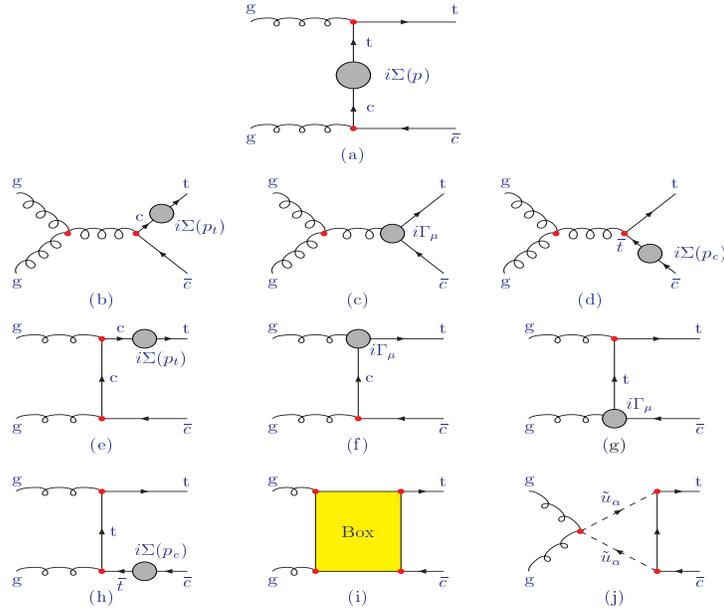,width=10cm, height=8.5cm}
\vspace*{-0.7cm}
\caption{Feynman diagrams for $g g \to t \bar{c}$. Additional
diagrams with the two gluons interchanged are not shown.}
\label{general}
\end{center}
\end{figure}

\begin{figure}[htb]
\begin{center}
\epsfig{file=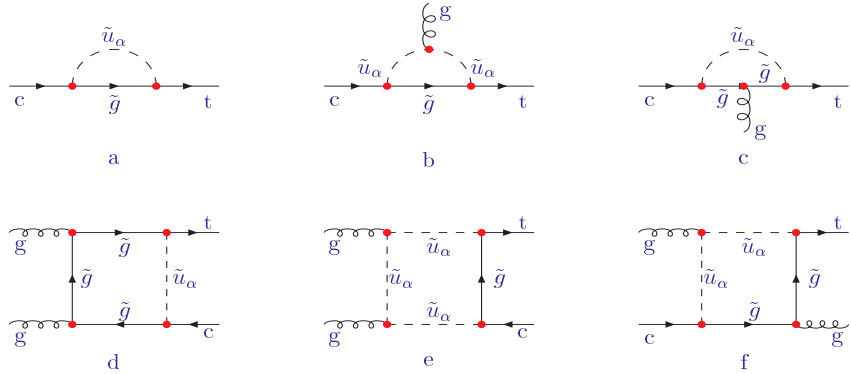,width=12cm, height=5.5cm }
\vspace*{-0.7cm}
\caption{SUSY-QCD contribution
to $c-t$ transition, $\bar{t} c g$ interaction and box diagrams
for $ g g \to t \bar{c}$.} \label{SQCD diagram}
\end{center}
\end{figure}

\begin{figure}[hbt]
\begin{center}
\epsfig{file=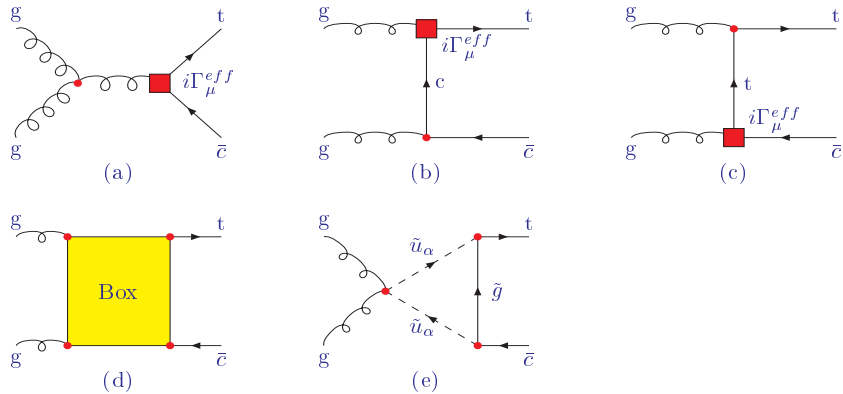,width=12cm, height=5.5cm}
\vspace*{-0.3cm}
\caption{Effective diagrams for the process $g g \to t \bar{c}$. Here
$\Gamma^{eff}_{\mu}$ is the effective $\bar{t}c g$ interaction
defined in Eq.~(\ref{eff}). } \label{effective}
\end{center}
\end{figure}

The key point of our method is the so-called ``effective vertex".
To illustrate this method we consider $ g g
\to t \bar{c}$ as an example. The Feynman diagrams for this process
are shown in Fig.~\ref{general}. The SUSY-QCD contributions
to the $c-t$ transition and the vertex $t\bar c g$,
as well as the box diagrams, are given in Fig.~\ref{SQCD diagram}.
Noting that the amplitude for Fig.~1(a) can be
split into two terms,  one containing a charm quark propagator,
and the other containing a top quark propagator
\begin{eqnarray}
M_a  \propto \frac{i} {\q_slash - m_t} i \Sigma(q)
\frac{i}{\q_slash - m_c} = \frac{i (\q_slash + m_t )}{m_c^2 -
m_t^2} \ i \Sigma(q) \ \frac{i}{\q_slash - m_c} +
 \frac{i}{\q_slash- m_t} \ i \Sigma (q) \  \frac{i (\q_slash + m_c)}{m_t^2 -
 m_c^2},   \label{technique}
\end{eqnarray}
we collect the first term together with Fig.~1(e, f),
and combine the second term together with
Fig.~1(g, h). After this arrangement, we can
define a momentum dependent effective $\bar{t}cg$ interaction as
\begin{eqnarray}
\Gamma^{eff}_{\mu} (p_t, p_c) &= & \Gamma_\mu^{\bar{t}cg}
(p_t,p_c) + i \Sigma (p_t) \  \frac{i (\p_slash_t + m_c)}{m_t^2 -
m_c^2} \Gamma_\mu^{\bar{q}qg} + \Gamma_\mu^{\bar{q}qg} \frac{i
(\p_slash_c + m_t )}{m_c^2 - m_t^2} \ i \Sigma(p_c), \label{eff}
\end{eqnarray}
where $\Gamma_\mu^{\bar{t}cg} $ is the penguin diagram
contribution to the effective interaction and
$\Gamma_\mu^{\bar{q}qg}$ is the usual QCD vertex. Then
the calculation of Fig.~\ref{general}(a-h) is
equivalent to the calculation of the ``tree" level transition
depicted in Fig.~\ref{effective}(a-c), which obviously has a
simpler structure. By following this method, our calculations
can be greatly simplified, since the effective  $\bar{t}cg$ interaction
appears in many processes, and the effective
$\bar{t}cg$ interaction is the same for all channels considered in the paper.
\vspace*{-.5cm}

Of course, the effective vertex $\bar{t}cg$ is model-dependent
since its components $\Gamma_\mu^{\bar{t}cg}$ and $\Sigma$ are
model-dependent. We obtained their expressions in SUSY QCD
analytically. We retain the tensor loop functions rather than
expanding them in terms of scalar loop functions as usual
\cite{Hooft}. This method makes our results quite compact and also
simplifies our Fortran codes which will be discussed below.

 The effective vertex $\bar{t}c g$ in
Eq.~(\ref{eff}) is a 4-component Lorentz vector and also
a $4 \times 4$ matrix in Dirac spinor space.  In its realization in
Fortran coding, we use a dimension-three array $V(i,j,k)$
with $i$ (=1,2,3,4) labeling the Lorentz index and  $j,k$ (=1,2,3,4)
labeling the spinor indices. We also use arrays to encode
other quantities such as Lorentz vectors, Dirac spinors, Lorentz tensors
and Dirac $\gamma$ matrices. The steps to calculate the effective
interaction in Fortran code can be then summarized as follows:
\begin{itemize}
\item  Input the matrices $P_{L,R}$, $\gamma^\mu P_{L,R}$ and
$\sigma^{\mu \nu}$.  For any other matrices
encountered in the calculation, we use $\gamma$ algebra to
generate its elements.

\item Use the mass splitting method\cite{Barger} to generate events
with momentums for the initial and final particles.

\item For a generated event with fixed  momenta,
      the components of any tensor loop function can be calculated numerically and stored in arrays.

\item  Generate the $\gamma $ matrices $\gamma^{\mu_1} \cdots
       \gamma^{\mu_n}$ and contract its Lorentz indices with those of
       tensor loop functions and those of quark momenta to obtain the
       effective vertex.
\end{itemize}
To calculate the amplitude of $g g \to t \bar{c}$, we also
need to calculate box diagrams. Such calculations are usually tedious
if the four-point  tensor loop functions are expanded in terms of
scalar loop functions.  Since we choose to retain the
tensor loop functions and contract the indices
numerically, our results are quite compact, as shown for
$ g g \to t \bar{c}$ in the Appendix.
The general expression for a box diagram is the sum of fermion chain
of the form $\left ( \bar{u} \gamma^{\mu_1} \cdots \gamma^{\mu_n} u
\right ) \times D_{\mu_i \cdots \mu_j} \times p_{\mu_k} \cdots $
and its numerical value is calculated by the following steps
\begin{itemize}
\item Input the wave functions for fermions and define the
multiplication of $\gamma$ matrix with the wave function.

\item Generate the tensor, say $ \bar{u} \cdots \gamma^{\mu_{n-1}}
\gamma^{\mu_{n}} u$, and contract its indices with those of loop
functions $D_{\mu_i \cdots \mu_j}$ and those of vectors involved,
to get the value of each term in the amplitude.
\end{itemize}

With the method introduced above, we can also easily calculate
other FCNC interactions. Let us take the calculation of $ c g \to t Z $ as
an example. Its Feynman diagrams can be obtained from
Figs.~\ref{general}-\ref{SQCD diagram} by removing those
involving triple-gluon interaction and gluon-gluino-gluino
interaction, and then replacing any gluon with Z boson.  Using the
technique from Eq.~(\ref{technique}) to introduce the effective
$\bar{t}c g$ interaction and the effective $ \bar{t}c Z$
interaction, one can again get simplified diagrams similar to
Fig.~\ref{effective}(b-e).

 Once  $ g g \to t \bar{c}$ is
calculated, evaluation of the others $ t \to c g$, $ c g \to t g$ and $ t \to c
g g$  becomes rather easy.  The decay  $t
\to c g$ is now a tree level interaction induced by the interaction
$\bar{t} c g$. The amplitudes (or their conjugates) for $ c g \to t
g $ and $ t \to c g g$ are related to that of $ g g \to t
\bar{c}$, and can be easily obtained by making some simple replacements
which can be easily realized in our code.

In the Appendix, we list the explicit forms of all the
penguin-induced FCNC interactions discussed in this paper. These
interactions are needed to obtain the effective top FCNC interactions.

\section{\bf Numerical results and dependence on SUSY parameters}

We know, from the discussion in Section II,
that the
calculations of the SUSY-QCD contributions induced by the flavor
mixings between left-handed scharm and stop, depend on the parameters
$M_{Q_{2,3}}$, $M_{U_3}$, $X_t = A_t - \mu \cot \beta$, $
m_{\tilde{g}}$, $\delta_{LL}$ and $\delta_{LR}$. In
this section, we investigate the dependence of the numerical
results on these parameters.  We first show the results for
top quark rare decays. These decay modes occur only via one
effective interaction and thus their dependence on the parameters
is relatively simple. We perform a comparative study and plot the
results for these decays together, to illustrate their dependence
on a given set of parameters. After analyzing the features of the
top quark rare decays, we extend the study to the FCNC top quark
productions. They usually involve two effective
interactions and several box diagrams, and thus their dependence
on SUSY parameters is more complex.

The SM parameters used in our calculations  are \cite{pdg}
\begin{eqnarray}
&& m_t = 172.7 {~\rm GeV}, ~~m_b= 4.8 {~\rm GeV}, ~~m_Z = 91.19 {~\rm GeV}, \nonumber \\
&& \sin \theta_W = 0.2228, ~~\alpha_s (m_t) = 0.1095, ~~\alpha =1/128. \label{sm-para}
\end{eqnarray}
After the assumptions discussed in Sec. III, about 10 SUSY
parameters are still involved.  We will show below the dependence
on SUSY parameters of the top FCNC processes.  When one of the
parameters is varied, the others will be fixed to their
``central'' values, taken as
\begin{eqnarray}
 \label{susy-para1}
M_{\rm SUSY}= M_{Q_{3}} = M_{U_3} = M_{Q_{2}}= 500 {~\rm GeV},
\quad X_t = 1000 {~\rm GeV}, \quad m_{\tilde{g}} = 250 {~\rm GeV},
\quad \tan \beta = 5.
\end{eqnarray}
The values of $\delta_{LL}$ and $\delta_{LR}$ will be shown in the
figures. With the exception of the last plot in this section (Fig.
12), we adopt the so-called $m_h^{\rm max}$ scenario \cite{mhmax}
which is widely discussed in Higgs physics, and which assumes that
all the soft mass parameters are degenerate
\begin{eqnarray}
M_{\rm SUSY} = M_{Q_i} = M_{U_i} = M_{D_i},
\end{eqnarray}
and that all the trilinear couplings are also degenerate,
$A_{u_i}=A_{d_i}$, with $X_t/M_{\rm SUSY}=2$.
In investigating the
processes $ t \to c h $ and $ c g \to t h $, we used the
loop-corrected lightest Higgs boson mass and the effective Higgs
mixing angle\cite{hollik,cao}. These two quantities involve two
additional parameters $\mu $ and $m_A$, which are fixed as
\begin{eqnarray} \label{susy-para2}
\mu = m_A = 500 {~\rm GeV}.
\end{eqnarray}

In our calculations we use CTEQ6L \cite{cteq} to generate the
parton distributions with renormalization scale $\mu_R $ and
factorization scale $\mu_F$, chosen to be $\mu_R = \mu_F = m_t$.
To make our predictions more realistic, we applied some kinematic
cuts. For example, for the three body decay $t \to c g g$  we
require that the energy of each decay product be larger than $15$
GeV and the separation of any two final states be more than $15^o$
in the top quark rest frame. For the top quark production
channels, we require that the transverse momentum of each produced
particle be larger than $15$ GeV and their pseudo rapidity be less
than 2.5 in the laboratory frame. Moreover, for $c g \to t$
followed by $t\to b W$, we do not require the top quark exactly on
mass shell and instead we require the invariant mass of bottom
quark and W boson in a region of $m_t - 3 \Gamma_t \leq M_{b W}
\leq m_t + 3 \Gamma_t $ ($\Gamma_t$ is the top quark width). This
requirement was once used in \cite{Hosch} to investigate the
observability of this channel at hadron colliders in the effective
Lagrangian framework.

Furthermore,  we vary the flavor mixings, $\delta_{LL}$ and
$\delta_{LR}$, over a wide range, with the only requirement that
they satisfy current collider searches for sparticles and Higgs
bosons\cite{pdg}:
\begin{eqnarray}
m_{\tilde{q}} \geq 96 {\rm ~GeV}, \quad m_{\tilde{g}} \geq 195  {\rm ~GeV}, \quad
m_h \geq 85  {\rm ~GeV}.  \label{bound5}
\end{eqnarray}
In principle, some low energy data, such as $b-s$ transition and
$\delta \rho$, can also constrain these mixings \cite{cao}. But
those so-called indirect constraints are usually quite
complicated. To simplify the discussion in this section we do not
impose these indirect constraints, but we address such a question
in the next section. We checked that the conclusions obtained in
this section  are valid in the region favored by these indirect
constraints \footnote{Another advantage of allowing the parameters
to vary within a large range is that, for most of the processes
considered in this paper, the dependence of their rates on
$\delta_{RL}$ and $\delta_{RR}$ is similar to that on
$\delta_{LR}$ and $\delta_{LL}$. But the indirect constraints on
them are quite different: while the indirect constraints on
$\delta_{LL}$ and $\delta_{LR}$ may be quite stringent, the limits
on $\delta_{RL}$ and $\delta_{RR}$ are rather weak \cite{cao}.
Therefore, the allowed range of $\delta_{RL}$ and $\delta_{RR}$ is
much larger than $\delta_{LL}$ and $\delta_{LR}$.}.

\begin{figure}[htb]
\begin{center}
\epsfig{file=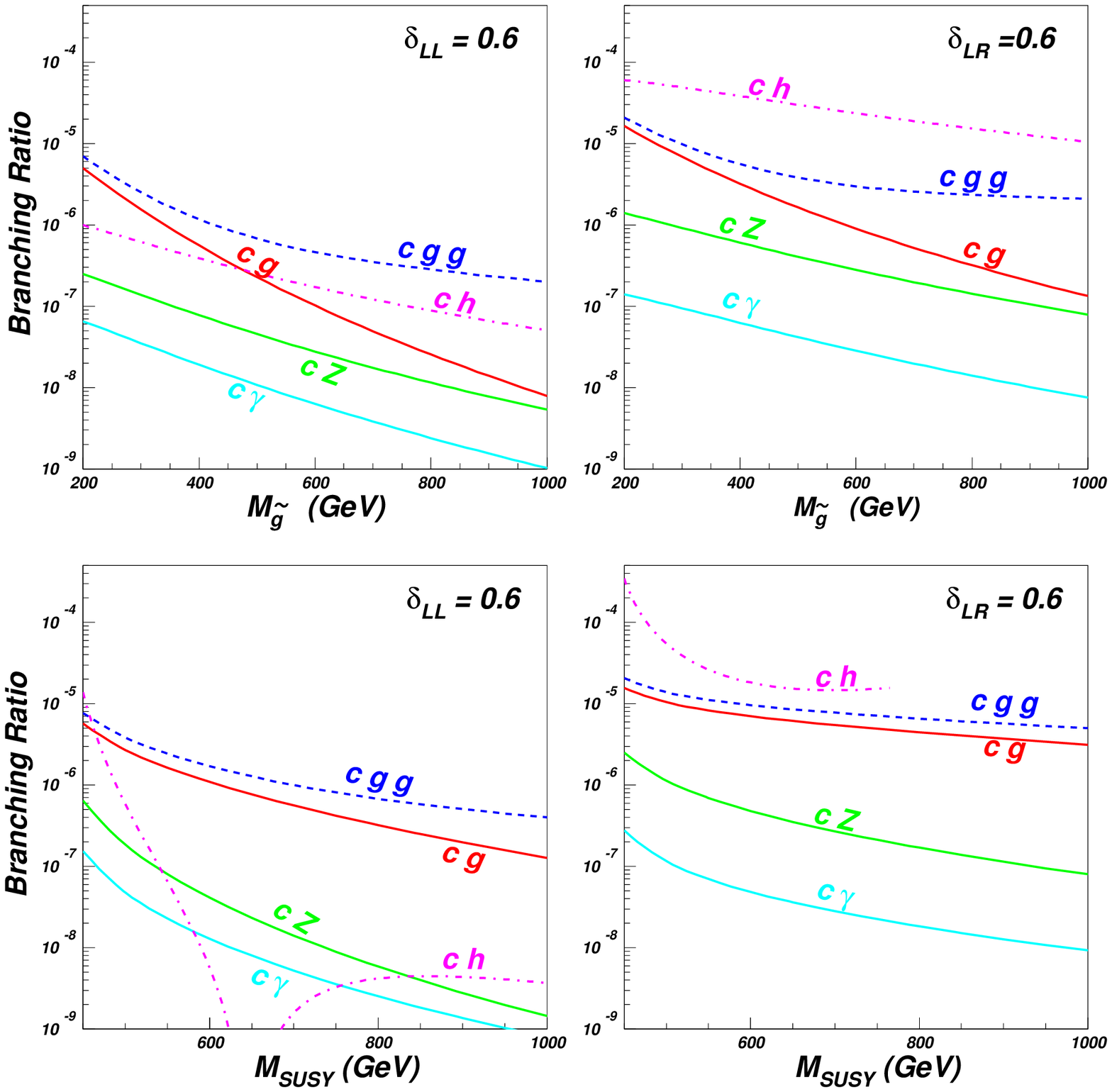,width=14cm,height=10cm}
\caption{The branching ratios of FCNC top quark decays. Unspecified parameters
are given in Eqs.~(\ref{sm-para}-\ref{susy-para2}). The
values of $\delta_{LL}$ and $\delta_{LR}$ are arbitrarily chosen,
and a smaller value will lower the rates, but not change the
tendencies of these curves.} \label{decay1}
\end{center}
\end{figure}

\begin{figure}[htb]
\begin{center}
\epsfig{file=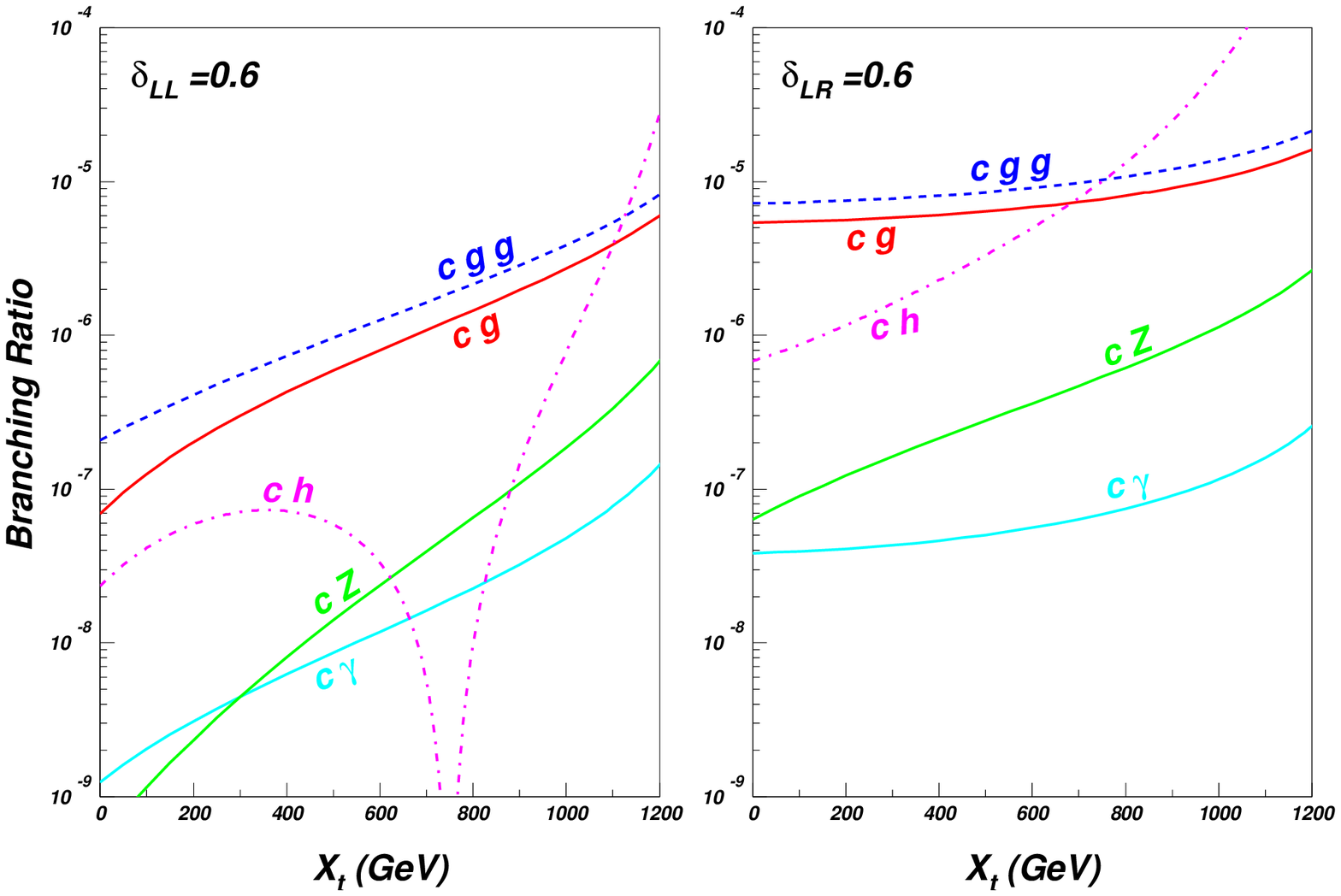,width=14cm,height=6cm } \caption{Same as
Fig.~\ref{decay1}, but as functions of $X_t (= A_t - \mu \cot \beta)$.}
\label{decay5}
\end{center}
\end{figure}

\begin{figure}[htb]
\begin{center}
\epsfig{file=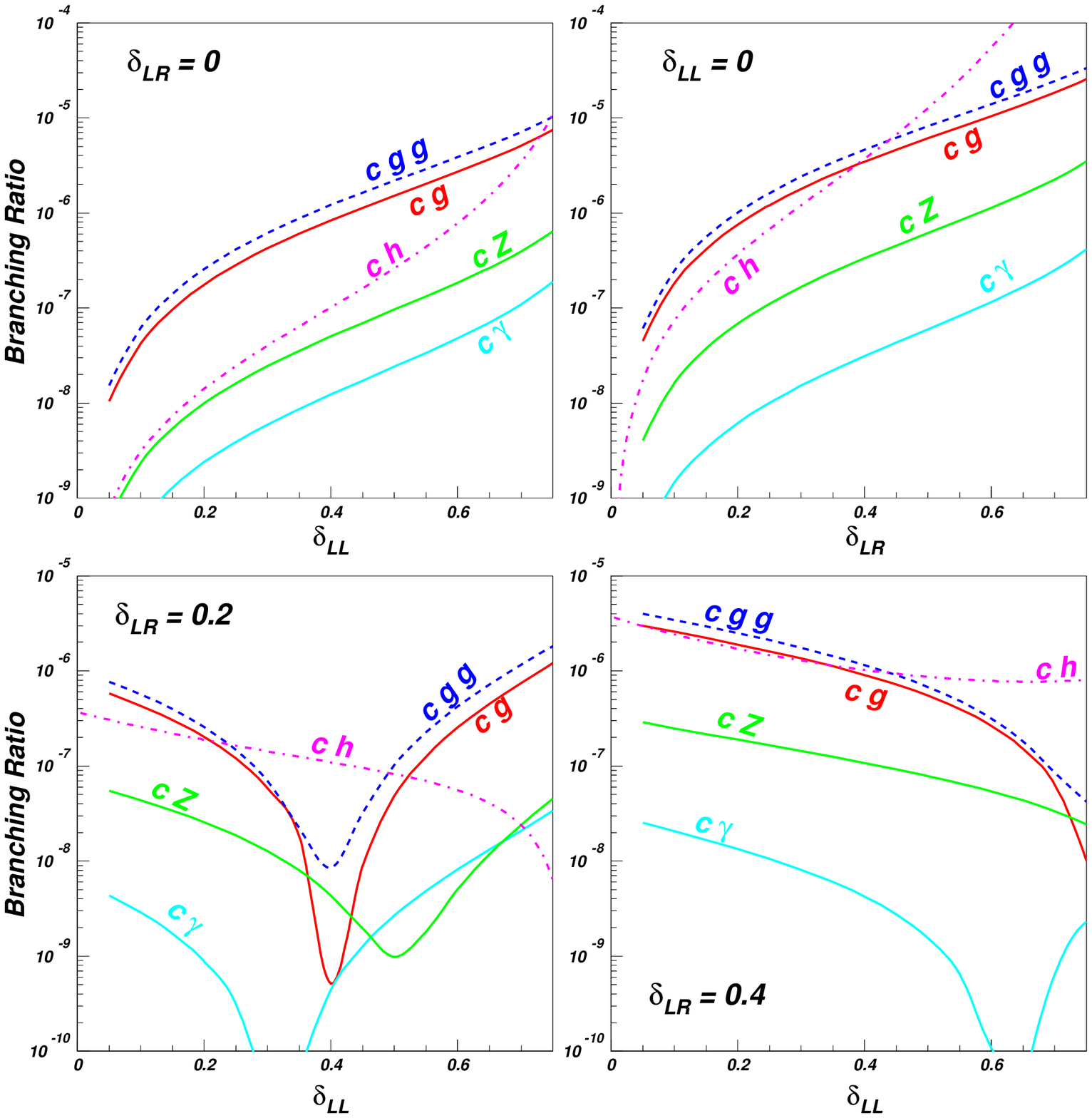,width=14cm,height=12cm } \caption{Same as
Fig.~\ref{decay1}, but as functions of the mixing parameters.}
\label{decay7}
\end{center}
\end{figure}

In Figs.~\ref{decay1}-\ref{decay7}, we present the branching ratios
of various FCNC top quark decays, defined  with respect to the width $\Gamma_t(t \to b W)$
($\simeq 1.45$ GeV). We plot the six branching ratios for the decays
in Eq.~(\ref{decy-1})-(\ref{fcncdecays}) as functions of
$m_{\tilde{g}}$, $M_{SUSY}$ and $X_t$ in the upper two diagrams of
Fig.~\ref{decay1}, the lower two diagrams of Fig.~\ref{decay1} and
Fig.~\ref{decay5}, respectively. We also show the dependence of the
branching ratios on the squark mixing parameter
$\delta_{LL}$ ($\delta_{LR}$) by
fixing the value of $\delta_{LR}$ ($\delta_{LL}$) in
Fig.~\ref{decay7}. These figures show some common features of
all the gauge boson decay modes. The first one is that, as the
sparticles become heavy, the branching
ratios drops monotonously (see Fig.~\ref{decay1}),
 a reflection of the decoupling property of
the MSSM. The second is that, as shown in Fig.~\ref{decay5}, the branching
ratios increase with the increase of $X_t$. This is because $X_t$
affects the squark mass splittings and could alleviate the cancellation between different loop
contributions. The third feature is that the branching
ratios increase rapidly
with the flavor mixing $\delta_{LL}$ and $\delta_{LR}$, because the flavor
mixings  can not only enhance the
coupling strength of squark flavor changing interactions, but also
enlarge the squark mass splittings. This
 is illustrated in the first two diagrams of
Fig.~\ref{decay7}. These last two effects combined
 make the branching ratios very sensitive to the flavor mixing
parameters.

By comparing the case $\delta_{LL} \neq 0 $ with the case
$\delta_{LR} \neq 0$ throughout Figs.~\ref{decay1}-\ref{decay7},
one finds that  $\delta_{LR}$ induces larger rates with weaker
dependence on sparticle masses. Furthermore, when both $\delta_{LL}$
and $\delta_{LR}$ are non-zero, as can be inferred from the last
two diagrams of Fig.~\ref{decay7}, the $\delta_{LL}$ and the
$\delta_{LR}$ dependences interfere destructively. To understand such
behaviors, we resort to the mass insertion method,  as it can give more intuitive results \cite{mass
insertion}. We take the decay $t \to c g$ as an example.
By gauge invariance, the general expression for
the effective $\bar{t}c g$ interaction takes the form
\begin{eqnarray}
\Gamma_\mu = F_1 (k^2) \bar{t}  T^a (k^2 \gamma_\mu - k_\mu
\k_slash ) P_L  c g^a -  m_t F_2 (k^2) \bar{t} T^a i \sigma_{\mu
\nu} k^\nu P_L c g^a \label{effective expression}
\end{eqnarray}
where $F_{1,2}(k^2)$ are form factors arising from loop
calculations. For the decay $ t \to c g$,  $F_1$ does not contribute,
 since the gluon momentum $k^\mu $ satisfies $k^2=0$
and $ k \cdot \epsilon = 0$. So only the dipole moment term is
relevant to our discussion.  Noting that the dipole changes both the flavor and the
chirality of the fermions, we may infer
the form of $m_t F_2$ in the mass insertion approximation. If only
$\delta_{LL}$ is considered for flavor changing, $m_t F_2 $ must
be
\begin{eqnarray}
m_t F_2 = \frac{m_t \delta_{LL}}{M_{SUSY}^2} A +
\frac{m_{\tilde{g}} \delta_{LL} m_t X_t}{M_{SUSY}^4} B,
\label{form1}
\end{eqnarray}
where $M_{SUSY} =max(m_{\tilde{g}}, M_{\tilde{q}})$, and $A, B$
are dimensionless constants coming from loop functions with
$1/M_{SUSY}^2$ factored out. The first term corresponds to the top
chirality
flipping contribution,  i.e., obtained by using the relation
$\bar{u}_t\!\p_slash_t = m_t \bar{u}_t$, while the second term corresponds to
the  $\tilde{t}_L-\tilde{t}_R$ mixing contribution, and  thus associated with
the gluino mass. The situation is quite different for $\delta_{LR}
\neq 0$, which alone can be responsible for both flavor
changing and chirality flipping. In this case, $m_t F_2$ should be
\begin{eqnarray}
m_t F_2 = \frac{m_{\tilde{g}} \delta_{LR}}{M_{SUSY}^2}C.
\label{form2}
\end{eqnarray}
where $C$, like $A$ and $B$, is a dimensionless constant coming
from loop functions with $1/M_{SUSY}^2$ factored out. Comparing
Eq.~(\ref{form1}) with Eq.~(\ref{form2}), we find that the
latter is larger if $m_{\tilde{g}} \gg m_t$. Assuming that $m_{\tilde{g}} \simeq M_{\tilde{q}}$, Eq.~(\ref{form1})
scales like $1/M_{SUSY}^2$ while Eq.~(\ref{form2}) scales like
$1/M_{SUSY}$. This explains the fact  that $\delta_{LR}$
induces larger rates with weaker dependence on sparticle masses.
Moreover, a detailed calculation shows that $m_t F_2$ in
Eq.~(\ref{form1}) is of opposite sign to that in Eq.~(\ref{form2}), which
 means that the $\delta_{LL}$ contribution tends to cancel the
$\delta_{LR}$ contribution.

The decay $ t \to c h$ has similar features to the decay into a
gauge boson except for a rather complicated dependence on
$M_{SUSY}$ and $X_t$, as shown in Figs.~\ref{decay1}-\ref{decay5}.
For the effective $\bar{t} c h$ interaction, with both the top and
the charm quarks on-shell, the general expression is
\begin{eqnarray}
\Gamma^{eff}_{\bar{t}ch} = D \bar{t} P_L c h,
\end{eqnarray}
where $D$ is a form factor emanating from the loop calculation.
This interaction,
like the effective $\bar{t}c g$ interaction, also involves both
flavor change and chirality flip. So in
some aspects, the behavior for $ t \to c h$ should be similar to $
t \to c g$. But its dependence on $X_t$ is more complicated than
other decay modes. The reason may be that $X_t$ not only affects
the masses and mixings of squarks, but also affects the Higgs
boson mass and its mixing angles, and, further, it enters the
interaction of $\tilde{q}^\ast \tilde{q} h$.

An interesting feature, shown in Figs.~\ref{decay1}-\ref{decay7} is
that the branching ratio of $ t\to c g$ is always smaller than
the higher order decay $ t \to c g g$. This was
observed in the SM \cite{tcgg-sm} and the MSSM
\cite{Gad-pptc-mssm}, and it indicates that the QCD corrections to
$t \to c g$ may be important. Two reasons may account for this behavior.
One is that the QCD factor in the amplitude-square for $t \to c g
g$ is much larger than that for $ t \to c g$. The other reason is that,
unlike the case $t \to c g$, $F_1$ in Eq.~(\ref{effective
expression}) also contributes to $t \to c g g$ and this
contribution is important. From
Figs.~\ref{decay1}-\ref{decay7}, one finds that, in most cases
\begin{eqnarray}
Br( t \to c g g ) > Br( t \to c g ) > Br ( t \to c Z ) > Br(t \to
c  \gamma),
\end{eqnarray}
and in some cases $Br( t \to c h) $ may be the largest.

\begin{figure}[htb]
\begin{center}
\epsfig{file=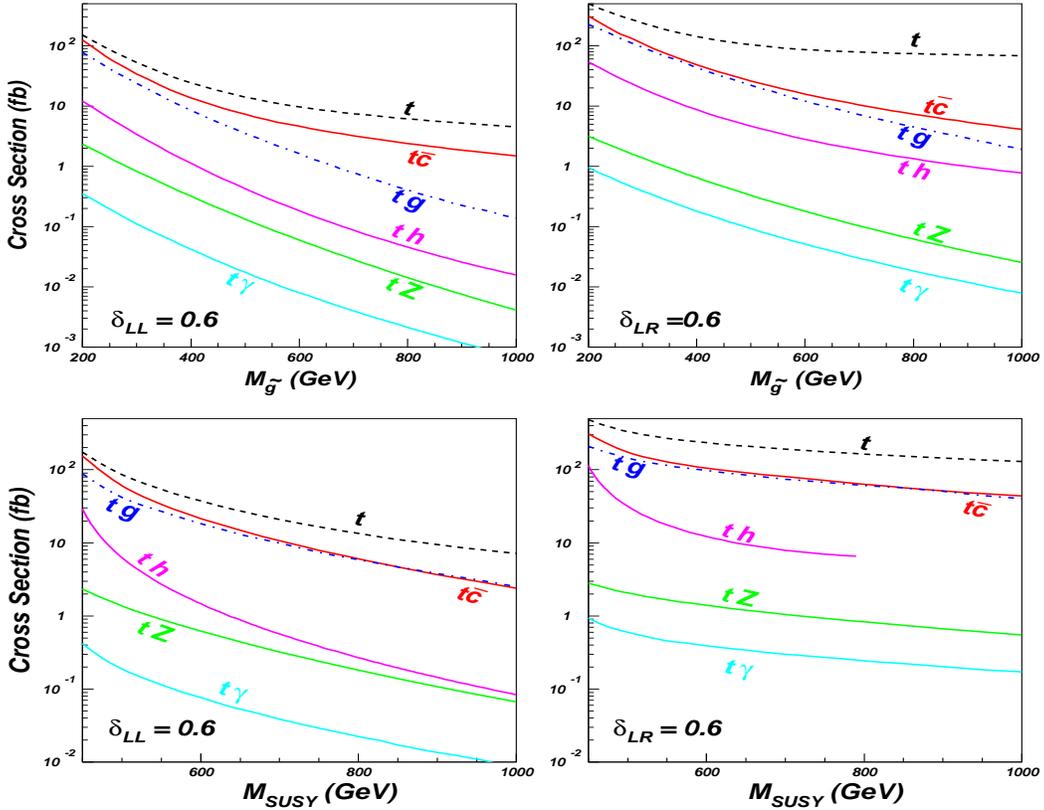,width=14cm,height=11cm }
\caption{The cross sections of the FCNC top quark productions at the LHC.
Each curve labeled by the final states corresponds to a certain production channel.}
\label{product1}
\end{center}
\end{figure}
\begin{figure}[htb]
\begin{center}
\epsfig{file=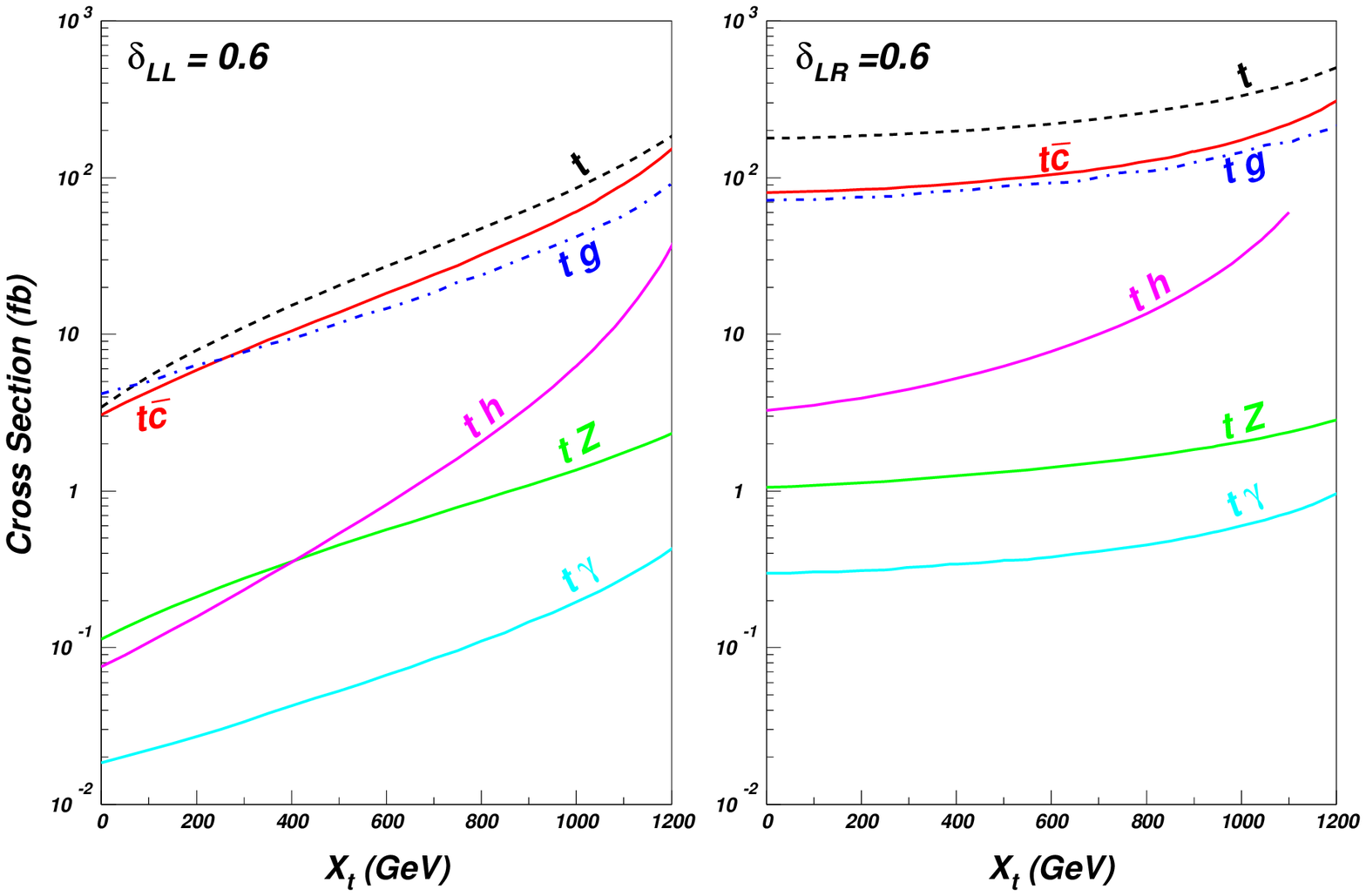,width=14cm,height=6cm }
\caption{Same as
Fig.~\ref{product1}, but as a function of  $X_t (= A_t - \mu \cot \beta)$.} \label{product5}
\end{center}
\end{figure}

\begin{figure}[htb]
\begin{center}
\epsfig{file=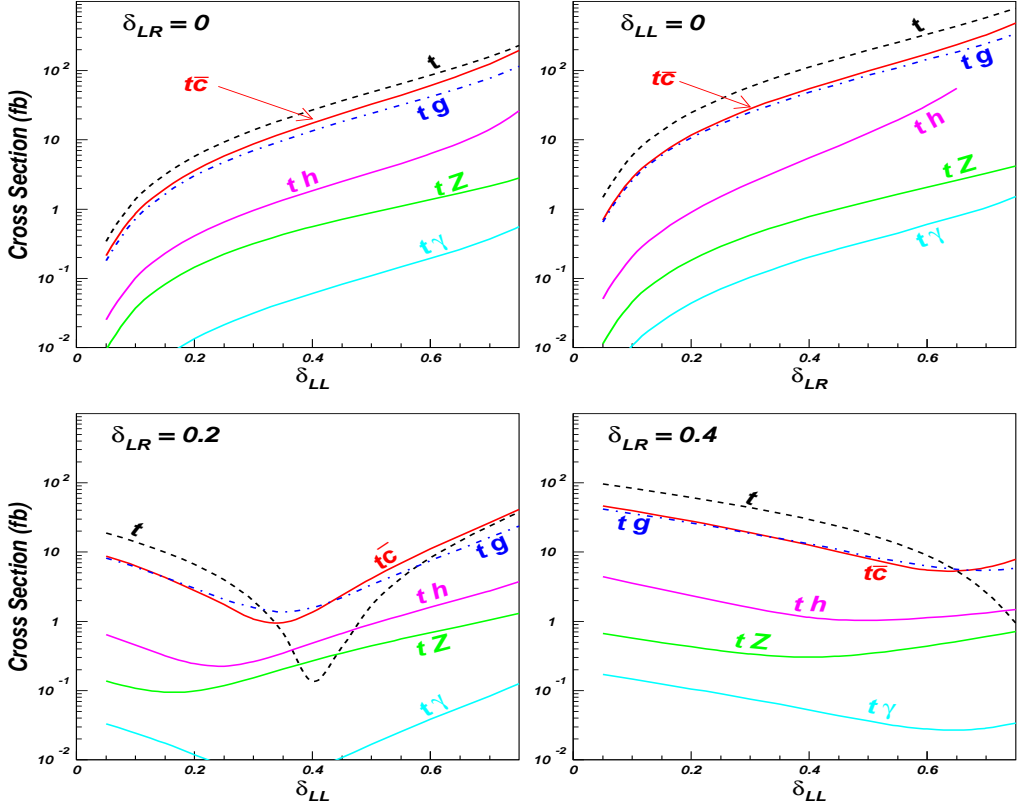, width=14cm,height=11cm }
\caption{Same as Fig.~\ref{product1}, but as a function of the mixing parameters.}
\label{product7}
\end{center}
\end{figure}

After giving the branching ratios for the FCNC top decays comparatively,
we now consider the FCNC top quark production channels. At the LHC the
production proceeds through the six parton-level processes shown in
Eqs.~(\ref{pro-1}-\ref{processes}), among which only $ g g \to t
\bar{c}$ has been extensively studied in the
MSSM\cite{pptc-mssm,Gad-pptc-mssm}. In calculating the
cross section for each channel, we also include the rate for its
charge conjugate channel. In the following discussions
we use the parton process to label its contribution to the hadronic
one.
Thus, the cross sections mentioned below refer to
the hadronic ones.

The dependence of the cross sections on SUSY parameters is plotted
in Figs.~\ref{product1}-\ref{product7}. These figures  exhibit the
same features as the FCNC top quark decays in
Fig.~\ref{decay1}-\ref{decay7}. Thus we do not repeat drawing
attention to the same features, and we only point out three
remarkable points of these diagrams. The first is that the rate of
$ c g \to t$ is generally larger than that for $ g g \to t
\bar{c}$. This is possible because the charm quark in the parton
distributions mainly comes from the splitting of a gluon
\cite{cteq} and thus $c g \to t $ can be seen as $ g g \to t
\bar{c}$ with the final charm quark going along the beam pipe.
Analyzing the signal versus background
\cite{Hosch,pptc-background}, it seems that $ c g \to t $ provides
a better opportunity for the observation of top production. The
second point is that the rate of $c g \to t g $ is comparable to
$g g \to t \bar{c}$ in most cases. This coincides with the results
in \cite{pptc-Han} where both modes were studied in the effective
Lagrangian approach. Since the two processes have the same signals
at colliders, namely single top plus one light-quark jet, one
should combine these two  when searching for the FCNC top
production events. The third point is that in most cases the cross
section for $c g \to t h$ is one order of magnitude smaller than
that for $g g \to t \bar{c}$. The reason is that although $c g \to
t h$ can proceed either through $\bar{t} c g$ interaction or  $
\bar{t} c h$ interaction, the two interactions interfere
destructively and thus the combined contributions are suppressed.

As shown above, in most of parameter space the
rates  are expected to be quite small at the LHC.  In
order to study the observability of such rare processes,
it is necessary to scan the whole
parameter space to figure out
the maximum value that each process can reach.
We call the `favorable region' the part of parameter
space which gives the maximum value for the rate of
an individual channel. Of course, such favorable region is
process-dependent. In what follows we take $g g \to t \bar{c}$ as
an example to carry out such a parameter scan explicitly. In
the next section, the maximal rates
for all processes will be tabulated, after discussing
the effects of the indirect constraints
on the flavor mixing parameters.

\begin{figure}[htb]
\begin{center}
\epsfig{file=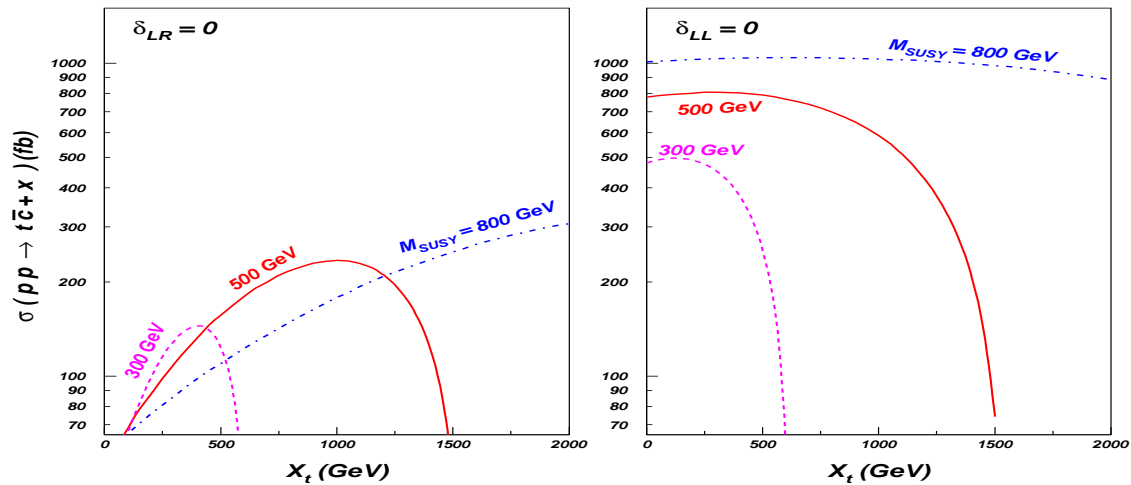,width=15cm,height=6.5cm} \vspace*{-0.3cm}
\caption{The maximal cross section of $t\bar c$ production at the
LHC, proceeding through the parton-level channel $g g \to t
\bar{c}$, as a function of $X_t$ by varying $m_{\tilde{g}}$ and
$\delta_{LL}$ or $\delta_{LR}$.} \label{m1}
\end{center}
\end{figure}

In Fig.~\ref{m1} we plot the maximal cross section
  for $ g g \to t \bar{c}$ as a function of $X_t$ with
non-zero $\delta_{LL}$ (left panel) and $\delta_{LR}$ (right panel).
The maximum value is obtained by fixing $M_{SUSY}$ and $X_t$ but varying
$m_{\tilde{g}}$ and $\delta_{LL}$ or $\delta_{LR}$.
In searching for such maximal values,
we required $m_{\tilde{q}} \geq 100$ GeV  and $m_{\tilde{g}} \geq 200$ GeV.

For non-zero $\delta_{LL}$, we find that the
parameter points for the maximal cross sections correspond to the
lightest squark mass and the gluino mass fixed at their
allowed lower bound, i.e., 100 GeV and 200 GeV, respectively. This
can be easily understood from the decoupling property of the MSSM.
At such optimum points, if $M_{SUSY}$ is fixed, $X_t$ value is
related to $\delta_{LL}$. Since both $X_t$ and
$\delta_{LL}$ are the off-diagonal elements in squark mass matrix,
a large $X_t$ will correspond to a small $\delta_{LL}$ and vice
versa. Then from Eq.~(\ref{form1}), one can
see that neither large nor small
$X_t$ can predict the largest rate for $ g g \to t \bar{c}$. This
explains the behavior of each curve in the left panel of
Fig.~\ref{m1}.

\begin{figure}[htb]
\begin{center}
\epsfig{file=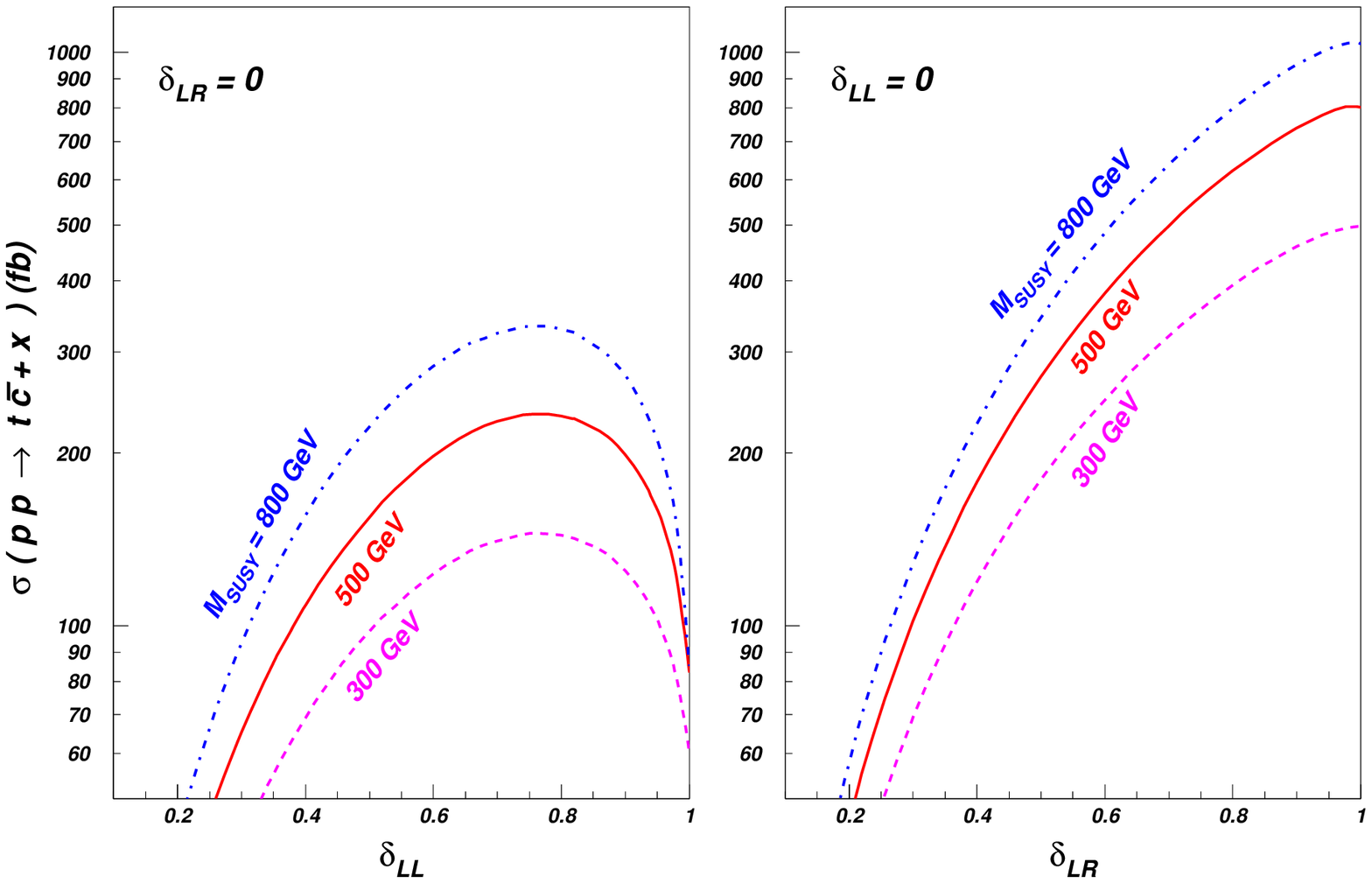,width=15cm,height=6.5cm } \vspace*{-0.3cm}
\caption{Same as Fig.~\ref{m1}, but as a function of mixing
parameters. } \label{m2}
\end{center}
\end{figure}

In Fig. ~\ref{m2}, we plot the maximal cross section as a function
of mixing parameters. For non-zero
$\delta_{LL}$ mixings (left panel), we see that the maximal value of each
curve lies at a moderate $\delta_{LL}$, which agrees with the
above analysis. Among the three curves for different $M_{SUSY}$,
the maximal value is obtained for $M_{SUSY} = 800 $ GeV. The reason is that
$M_{SUSY} = 800 $ GeV implies heavier masses for the other two squarks
and alleviates the cancellation between different diagram
contributions.
\begin{figure}[htb]
\begin{center}
\epsfig{file=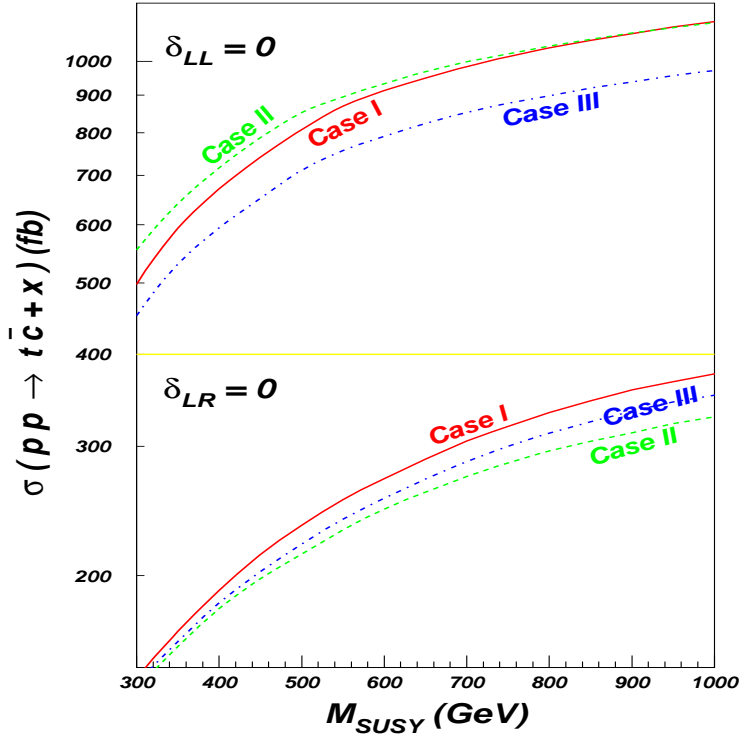,width=10cm,height=10cm }
\caption{Same as Fig.~\ref{m1}, but as a function of $M_{SUSY}$ under
different cases specified in the text.}
\label{m5}
\end{center}
\end{figure}

For the case of non-zero $\delta_{LR}$, maximal cross sections
also appear  at the lower bounds of the lightest squark mass and
the gluino mass. This property together with the effective
$\bar{t} c g$ coupling in Eq.~(\ref{form2}), imply that the
maximal value for each curve should lie at a large $\delta_{LR}$,
or by the correlation, at small $X_t$. (Both $X_t$ and
$\delta_{LR}$ appear as the off-diagonal elements in squark mass
matrix. Thus, for a given $M_{SUSY}$, their large values enlarge
the mass splitting between squark mass eigenstates and lead to a
small mass for the lightest squark. Due to the lower bound on the
lightest squark mass, $\delta_{LR}$ and $X_t$ cannot be both
large.) This is in agreement with the behaviors of the curves for
non-zero $\delta_{LR}$ shown in the right panels of
Figs.~\ref{m1}-\ref{m2}.  For the right panel of Fig.~\ref{m1}, we
checked that for $M_{SUSY} = 800 $ GeV, varying $X_t$ from 0 to
$500$ GeV only resulted in a change of $0.006$ for $\delta_{LR}$.
This is the reason that the maximal cross section is insensitive
to $X_t$ for $X_t \leq 500$ GeV.

In Fig.~\ref{m5} we show the dependence of the maximal cross
section on $M_{SUSY}$ for non-zero $\delta_{LL}$ or $\delta_{LR}$ values.
To get the maximal cross sections, we fix the lightest squark mass
as $100$ GeV and vary the value of $\delta_{LL}$ or $\delta_{LR}$.
We considered three cases,
\begin{itemize}
\item Case I: $M_{Q_2} = M_{Q_3} = M_{U_3} = M_{SUSY}$, \item Case
II:  $M_{Q_3} = M_{U_3} = M_{SUSY} $, $M_{Q_2} = 1.2 M_{SUSY}$,
\item Case III: $M_{Q_3} = M_{SUSY}, M_{U_3} = 0.8 M_{SUSY},
M_{Q_2} = 1.2 M_{SUSY}$.
\end{itemize}
These cases are motivated by mSUGRA model
\cite{sugra} where the three squark masses are generated from the same soft
breaking mass parameter $m_0$ at supersymmetry breaking scale, but are split
due to quark Yukawa couplings when they evolve down to electroweak
scale \cite{spectrum}. From this figure we see that the maximal
cross section values increases with $M_{SUSY}$, and for non-zero
$\delta_{LL}$ ($\delta_{LR}$), case I (II) gives the largest
prediction for the cross sections. These results indicate that in
searching for the
maximum values of the cross section, we should treat $M_{Q_2}$,
$M_{Q_3}$ and $M_{U_3}$ as free parameters and explore all the
possibilities for  these masses.

Note that in the above discussions we have only considered the
case for non-zero $\delta_{LL}$ or $\delta_{LR}$. But as pointed
below Eq.~(\ref{decompose}), most of our conclusions should  be
valid for the case non-zero $\delta_{RL}$ or $\delta_{RR} \neq 0$
with an interchange of $L$ and $R$.

\section{\bf Low-energy constraints on FCNC top quark interactions}
\subsection{Constraints on scharm-stop flavor mixings}
We know, from the discussions in the preceding section, that the
rates of the FCNC top quark processes depend strongly on the
flavor mixing parameters $\delta_{LL}$ and $\delta_{LR}$, which
are treated as free. Of course, such a treatment is
informative and useful to the LHC experiments since it allows to
directly place limits on the mixing parameters once the
measurements are made at the LHC.

However, it is worth noting that these mixing parameters may be subject to
various direct and indirect experimental constraints. Firstly,
since the mixing terms appear as the non-diagonal elements of
squark mass matrices, they can affect the squark mass spectrum,
especially by enlarging the mass splitting between squarks. Therefore they
should be constrained by the squark mass bounds from the direct
experimental searches. At the same time, since the squark loops
affect the precision electroweak quantities such as $M_W$ and the
effective weak mixing angle
$\sin^2\theta_{eff}$\cite{hollik,mw-mssm}, such mixings could also be
constrained by precision electroweak measurements. As
shown in\cite{hollik,cao}, to a good approximation, the
supersymmetric corrections to the electroweak quantities contribute
through the $\delta\rho$ parameter and thus sensitive to the mass
splitting of squarks. Secondly, the processes governed by $b \to s$
transition like $B_s-\bar B_s$ mixings \cite{Ball} and $b \to s
\gamma$\cite{bsr} can provide rich information about the $\tilde
s-\tilde b$ mixings. Through the SU(2) relation between the up-squark
and down-squark mass matrices (see Eq.~(\ref{SU2})) and also
through the electroweak quantities (since all squarks contribute
to electroweak quantities via loops), the information can be
reflected in the up-squark sector and hence constrain the scharm-stop
mixings. Thirdly, we note that the chiral flipping mixings of
scharm-stop come from the trilinear $H_2 \tilde{c}_L^\ast
\tilde{t}_R$ interactions\cite{susyflavor}. Such interactions can
lower the lightest Higgs boson mass $m_h$ via squark loops and
thus should be subject to the current experimental bound on $m_h$.

In \cite{cao} these constraints were examined and it was shown that,
 although they usually depend on additional unknown parameters
in the down-type squark mass matrix, a combined analysis can still
severely restrict the mixings $\delta_{LL}$ and $\delta_{LR}$
in most cases. Here we extend the analysis of  \cite{cao}
by providing more examples about these constraints and then
perform a detailed investigation of the maximum rates for various
top quark FCNC processes both with and without these constraints.

\begin{figure}[htb]
\begin{center}
\epsfig{file=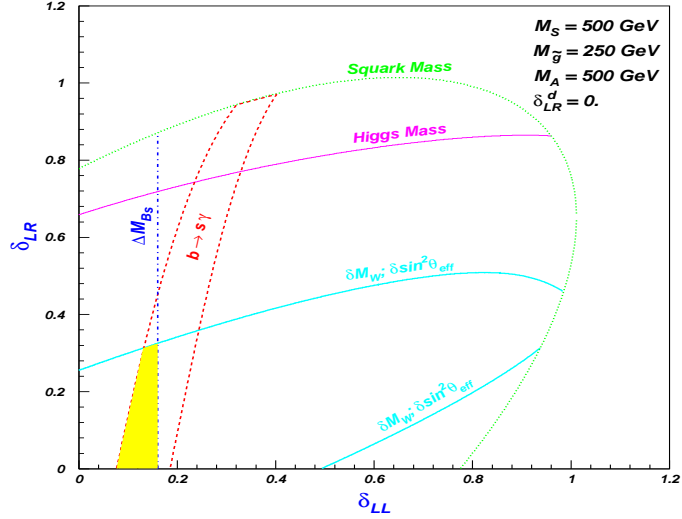,width=9.cm,height=7.cm } \vspace*{-0.3cm}
\caption{Experimental constraints on $\delta_{LL}$ versus
$\delta_{LR}$ with the parameters given in Eq.~(\ref{susy-para}).
The region under or left to each curve corresponds to the allowed
region. The dashed-line enclosed area is allowed by $b \to s
\gamma$. The colored area is the overlap region satisfying all the
constraints.} \label{constr1}
\end{center}
\end{figure}

The constraints considered in our paper are $b \to s\gamma$,
$B_s-\bar{B}_s$ mixing, $\delta M_W $ and $\delta \sin^2
\theta_{eff}$ and the lightest Higgs boson mass. In the following,
we first recapitulate these constraints and then apply them
to scharm-stop mixings.
\begin{itemize}
\item[(1)] {\it $ b \to s \gamma$:} In the MSSM, there are four kinds of
loops contributing to $ b \to s \gamma$ mediated respectively by
the charged Higgs bosons, charginos, neutralinos and gluinos, each
of which may be sizable \cite{bsr}. For a light charged Higgs
mass, the contribution from the charged Higgs is quite large and
always has the same sign as the SM contribution, and thus
enhance the branching ratio to very high values \cite{charged higgs}. The
current $b \to s \gamma$ data require either a  sufficiently heavy
charged Higgs boson, or its contribution should  be canceled by other parts
of SUSY effects. For the other three kinds of contributions, depending
on SUSY parameters, they may interfere constructively or
destructively with the SM effects and thus their relative sizes are not
fixed \cite{bsr-character}. The effect of these properties of SUSY
 on $ b \to s \gamma$ makes it necessary to consider all the
contributions simultaneously in discussing $ b \to s \gamma $
constraints on the squark flavor mixing parameters.

Current measurement of the branching ratio for $ b \to s \gamma$
is rather precise, with $3 \sigma$ bounds given by \cite{heavy
flavor}
\begin{eqnarray}
2.53 \times 10^{-4} < Br ( b \to s \gamma ) < 4.34 \times 10^{-4}.
\label{bound1}
\end{eqnarray}
With the SM prediction $Br^{NLO}( b \to s \gamma ) = (3.53 \pm 0.30)
\times 10^{-4}$ \cite{bsr-sm} and a favored negative $C_7$ by $ b
\to s l^+ l^- $ \cite{bsll}, where $C_7$ denotes the Wilson
coefficient for the electromagnetic dipole operator ${\cal O}_7$,
this decay can severely restrict the SUSY parameters. Our numerical
results indicate that it is very sensitive to $\delta^d_{LR}$ and
$\delta^d_{RL}$ for most cases, and for large $\tan \beta$, it is
sensitive to $\delta_{LL}$ as well. The same results also
indicate that $b \to s \gamma$ depends weakly on $\delta_{LR}$ which
affects the decay via chargino-mediated loops,  but under all circumstances $ b
\to s \gamma$ is not sensitive to $\delta_{RL}$ and $\delta_{RR}$.

\item[(2)] {\it $ B_s-\bar{B}_s$ mixing:} In the MSSM, although the loops
mediated by charged Higgs boson, chargino and neutralino
contribute to $B_s-\bar{B}_s$ mixing, their effects are generally
much smaller than the SM contribution\cite{Ball}. So when
discussing the constraint of $B_s-\bar{B}_s$ mixing on squark flavor
mixing parameters, we only consider gluino contributions.
Recently, the D0 collaboration gave the first two-side bound
on the mass splitting between $B_s$ and $\bar{B}_s$ \cite{Abazov:2006dm}
\begin{eqnarray}
17 ~{\rm ps}^{-1} < \Delta M_s < 21  ~{\rm ps}^{-1}   \quad (90\%~  {\rm C.L.}) \label{B_s}
\end{eqnarray}
This result is in agreement with the SM prediction, which is
estimated as $21.3 \pm 2.6 ~{\rm ps}^{-1}$ by the UTfit group
\cite{Bona:2005vz} and $20.9^{+4.5}_{-4.2} ~{\rm ps}^{-1}$ by the
CKMfitter group \cite{Charles:2004jd}. After considering various
uncertainties, the bounds in Eq.~(\ref{B_s}) can be re-expressed
as\cite{Endo:2006dm}
\begin{eqnarray}
0.55 < | 1 + M_{12}^{SUSY}/M_{12}^{SM} | < 1.37,   \label{bound2}
\end{eqnarray}
where $M_{12}$ is the transition matrix element for
$B_s-\bar{B}_s$ transition. As pointed out in\cite{Ball},
$B_s-\bar{B}_s$ mixing is very sensitive to the combinations
$\delta_{LL} \delta_{RR}^d $ and $\delta_{LR}^d \delta_{RL}^d$
and thus can put rather stringent constraints on any of $\delta^d$s.

\item[(3)] {\it $\delta M_W $ and $\delta \sin^2 \theta_{eff}$:} In the MSSM,
the corrections to $M_W$ and $\sin^2 \theta_{eff} $ involve the
calculation of the gauge boson self energy, and among all kinds of
contribution to the self energy, those from squark loops are most
important\cite{rho}.  As a good approximation, $\delta M_W$ and
$\delta \sin^2 \theta_{eff}$ are related to $\delta \rho$
by\cite{hollik}
\begin{eqnarray}
\delta M_W \simeq \frac{M_W}{2} \frac{c_W^2}{c_W^2 -s_W^2} \delta \rho, \nonumber \\
\delta \sin^2 \theta_{eff} \simeq -\frac{c_W^2 s_W^2}{c_W^2
-s_W^2} \delta \rho ,  \label{relation}
\end{eqnarray}
where
\begin{eqnarray}
 \delta \rho \equiv \frac{\Sigma_Z(0)}{M_Z^2} - \frac{\Sigma_W(0)}{M_W^2}.
\end{eqnarray}
Since the couplings are stronger for left-handed squarks
than for right-handed squarks, $\delta \rho $ is sensitive to the
mass splittings between left-handed up-squarks and down-squarks \cite{rho}.
As far as $\delta_{LL}$ is concerned, due to
the $SU(2)$ relation in Eq.~(\ref{SU2}), it changes up-squark and
down-squark mass spectra simultaneously and thus its effects on
$\delta \rho $ are generally small even for large $\delta_{LL}$.
For $\delta_{LR}$ and $\delta_{RL}$, they are independent of
$\delta_{LR}^d $ and $ \delta_{RL}^d$, which are very small as
required by $b-s$ transition \cite{bsr,Ball,B-summary}. Thus
a large $\delta_{LR}$ or $\delta_{RL}$ can induce a sizable mismatch
between up-squark and down-squark mass spectra. As a result, large
$\delta_{LR}$ or $\delta_{RL}$ can significantly change $\delta \rho$
\cite{cao}.

With the recent analysis of the LEP data, the uncertainties in
measuring $M_W$ and $\sin^2 \theta_{eff}$ were significantly
lowered to read \cite{lep}
\begin{eqnarray}
\delta M_W < 34 {\rm ~MeV}, \quad \delta \sin^2 \theta_{eff} < 15 \times
10^{-5}. \label{bound3}
\end{eqnarray}
These uncertainties imply that the new physics influence on
$\delta \rho $ should be lower than $5.5 \times 10^{-4}$.

\item[(4)] {\it Higgs boson mass $m_h$:} In the MSSM the
loop-corrected lightest
Higgs boson mass $m_h$ is defined as the pole of the corrected
propagator matrix, which can be obtained by solving the equation
\cite{Dabelstein}
\begin{eqnarray}
 & & \left[p^2 - m_{h, tree}^2 + \hat{\Sigma}_{hh}(p^2) \right]
 \left[p^2 - m_{H, tree}^2 + \hat{\Sigma}_{HH}(p^2) \right]
-\left[\hat{\Sigma}_{hH}(p^2)\right]^2 = 0 , \label{mass-equation}
\end{eqnarray}
where $m_{h, tree}$ and $m_{H, tree}$ are the tree-level masses of
the neutral Higgs bosons $h$ and $H$, and  $\hat\Sigma_i (p^2)$ ($i= h h$, $h H$, $H H$)
are the renormalized Higgs boson self energies. Among all SUSY
contributions to the Higgs boson self energies, those from top and
stop loops are by far dominant because of the large top quark Yukawa
couplings\cite{h-original}. In the presence of the flavor mixings
in the up-squark mass matrix, stops will mix with other squarks,
and in this case, the dominant contribution comes from the up-squark
sector \cite{hollik,feynhiggs}.

Our results indicate that the Higgs boson mass is more sensitive
to $\delta_{LR}$ than to $\delta_{LL}$.  The reason is that
$\delta_{LL}$ affects the Higgs boson mass only by changing the
squark interaction through the unitary matrix $\Gamma$ in
Eq.~(\ref{interaction}) while $\delta_{LR}$ can
also appear directly in the coupling of trilinear $H_2 \tilde{c}_L
\tilde{t}_R$ interaction, since the $\delta_{LR} M_{Q_2} M_{U_3}$ term
in up-squark mass matrix comes from this trilinear
interaction, and this  can reduce the lightest Higgs
boson mass.
Current collider searches for a MSSM Higgs boson have given the
lower bounds on $m_h$ in five benchmark scenarios in Higgs
physics\cite{benchmark}. In our calculation, we use the limit
\begin{eqnarray}
m_h >\left \{
\begin{array}{ll} 92.8  {~\rm GeV} & \quad {~\rm for} \ m_h^{max}\ {~\rm scenario} \\
                  & \\
85   {~\rm GeV}  & \quad   {~\rm for ~other ~scenarios}  \end{array} \right . \label{bound4}
\end{eqnarray}
\end{itemize}
As the first example of these constraints, we consider the
$m_h^{max}$ scenario with the parameters given in
Eqs.~(\ref{susy-para1},\ref{susy-para2}). To determine these
constraints, we also need to specify the values of the parameters
in the down-squark mass matrix and the gaugino mass $M_2$. In
Fig.~\ref{constr1} we show the allowed region in
$\delta_{LL}-\delta_{LR}$ plane for the parameters
\begin{eqnarray}
&&M_{S}= M_{Q_{i}} = M_{U_i} = M_{D_{i}}= 500 {~\rm GeV}, ~~X_t 1000 {~\rm GeV}, ~~M_{\tilde{g}} = 250 {~\rm GeV}, ~~\tan \beta
=5,  \nonumber \\
&&\mu = M_A = M_2 = 500 {~\rm GeV},~~ \delta_{RL}=\delta_{RR}0,~~\delta^d_{LR}=\delta^d_{RL}=\delta^d_{RR}=0. \label{susy-para}
\end{eqnarray}

\begin{figure}[htb]
\begin{center}
\vspace*{-0.3cm} \epsfig{file=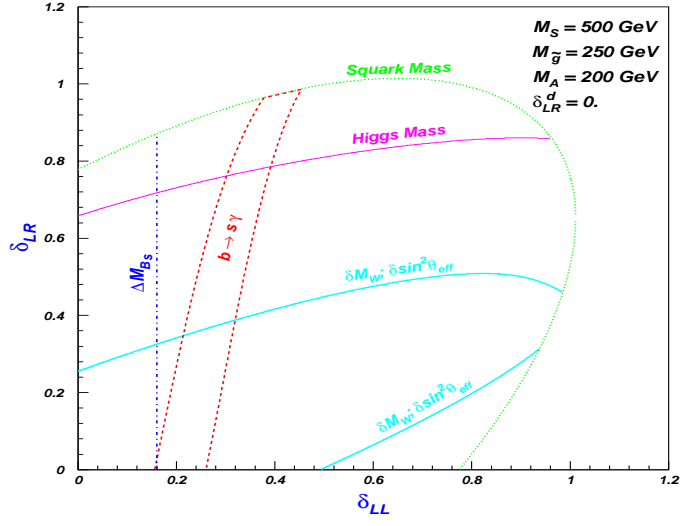,width=9.0cm,height=7.cm }
\caption{Same as Fig.~\ref{constr1}, but for $M_A = 200 $ GeV.}
\label{constr2}
\end{center}
\end{figure}

\begin{figure}[htb]
\begin{center}
\epsfig{file=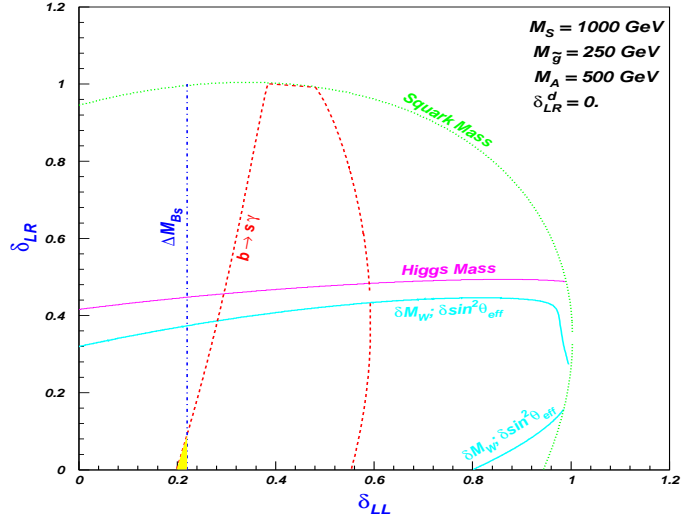,width=9.0cm,height=7cm}
\vspace*{-0.3cm}
\caption{Same as Fig.~\ref{constr1}, but for $ M_S= 1 $ TeV.}
\label{constr3}
\end{center}
\end{figure}

\begin{figure}[htb]
\begin{center}
\epsfig{file=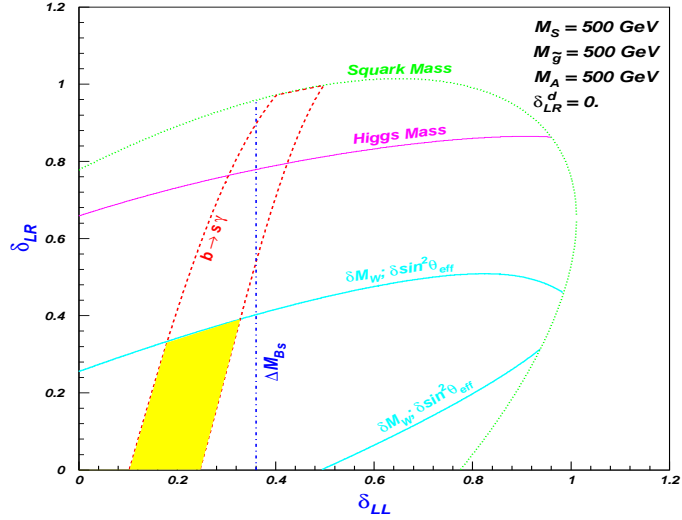,width=9.0cm,height=7cm}
\vspace*{-0.3cm}
\caption{Same as Fig.~\ref{constr1}, but for $M_{\tilde{g}} = 500 $ GeV.}
\label{constr4}
\end{center}
\end{figure}

\begin{figure}[htb]
\begin{center}
\epsfig{file=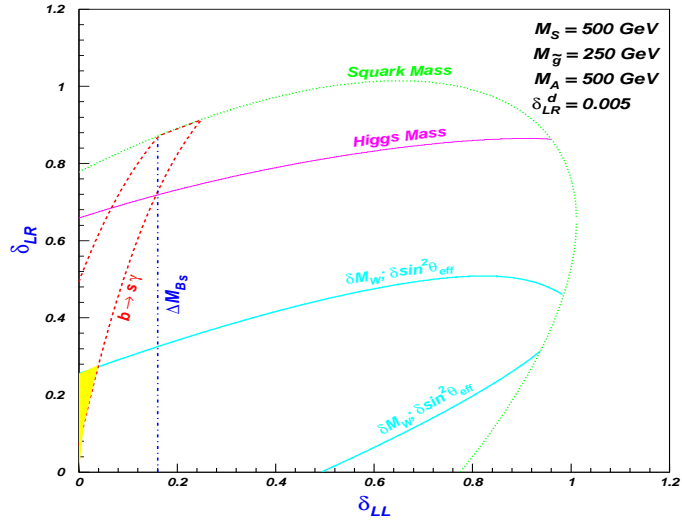,width=9.0cm,height=7cm}
\vspace*{-0.3cm}
\caption{Same as Fig.~\ref{constr1}, but for $\delta^d_{LR}=0.005$.}
\label{constr5}
\end{center}
\end{figure}

\begin{figure}[htb]
\begin{center}
\epsfig{file=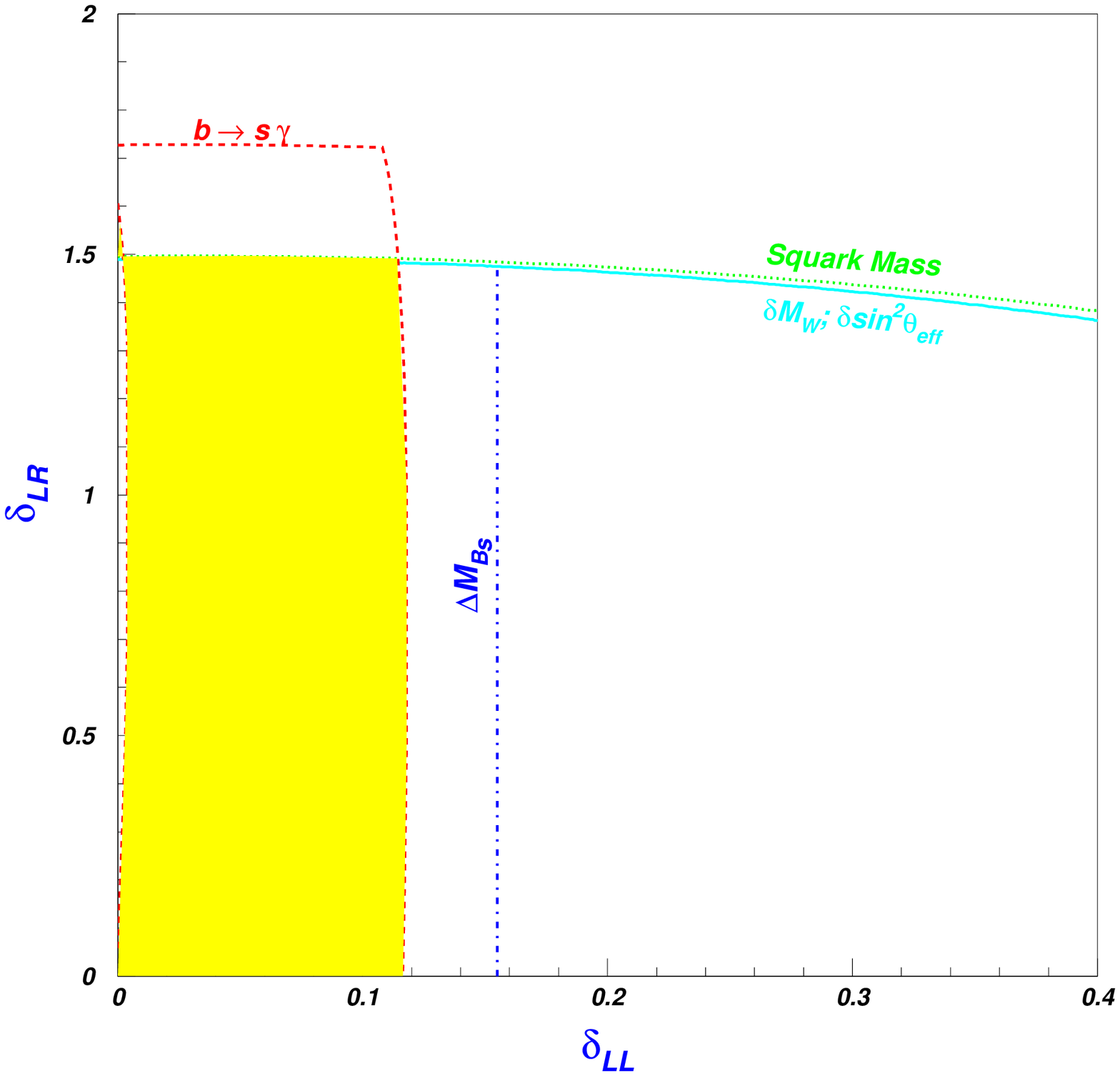,width=9.0cm,height=7cm}
\vspace*{-0.3cm}
\caption{Same as Fig.~\ref{constr1}, but for the parameters in
Eq.~(\ref{max-para}).
In this case the Higgs boson mass cannot impose any constraints.}
\label{constr6}
\end{center}
\end{figure}

Each curve in this figure corresponds to an experimental bound
from Eqs.~(\ref{bound5}), (\ref{bound1}), (\ref{bound2}),
(\ref{bound3}) and (\ref{bound4}); while the colored area is the
overlap region satisfying all the constraints. The distinctive
character of this figure is that $ b \to s \gamma $ requires a
non-zero $\delta_{LL}$. The reason is that with the fixed
parameters in Eq.~(\ref{susy-para}), especially with
$\delta^d_{LR}= 0 $, the charged Higgs contribution enhances the
SM contribution and, as a result, a none-zero gluino contribution
is needed to cancel such effects. From Fig.~\ref{constr1} we see
that the allowed region is mainly determined by $ b \to s \gamma$
and $ B_s-\bar{B}_s$ mixing. To show the dependence of such
allowed region on SUSY parameters, we vary the values of $M_A$,
$M_S$, $M_{\tilde{g}}$ and $\delta_{LR}^d$ one at a time, and get
the allowed region (colored area) in Figs.~\ref{constr2},
\ref{constr3}, \ref{constr4} and \ref{constr5}, respectively.
Explicitly, Fig.~\ref{constr2} corresponds to the parameters in
Eq.~(\ref{susy-para}) but with $M_A = 200 $ GeV,
Fig.~\ref{constr3} corresponds to the parameters in
Eq.~(\ref{susy-para}) but with $M_S = 1000 $ GeV ,  and
Fig.~\ref{constr4} and Fig.~\ref{constr5} are drawn in a similar
way. The shift of the allowed region can be well understood by the
properties of each constraint. Take Fig.~\ref{constr2} as an
example. As $M_A$ becomes smaller, the charged Higgs contribution
to $b \to s \gamma$ further enhance the SM contribution and
consequently a larger $\delta_{LL}$ is needed to cancel such
effect and satisfy the bound in Eq.~(\ref{bound1}). Since the
constraint from $B_s-\bar{B}_s$ mixing is not changed, the overlap
region diminishes gradually and finally vanishes for $M_A \simeq
200 $ GeV.

As the second example, we consider the following parameters as an
input
\begin{eqnarray}
&&M_{Q_2}= 400  {\rm ~GeV}, \quad  M_{Q_3} = 1000  {\rm ~GeV},
\quad M_{U_3} = 120  {\rm ~GeV}, \quad M_{\tilde{g}} = 196  {\rm
~GeV}, \quad M_A = 160  {\rm ~GeV},
\nonumber \\
&&X_t = 33  {\rm ~GeV}, \quad \mu = -330  {\rm ~GeV}, \quad
M_2 = 860  {\rm ~GeV}, \quad M_{D_i} = M_{Q_1}= M_{D_1} = 500  {\rm ~GeV}, \nonumber \\
&& \delta_{LR}^d= 0.0026, \quad \delta_{RL}=\delta_{LR}= 0, \quad
\delta_{RL}^d=\delta_{RR}^d=0.  \label{max-para}
\end{eqnarray}
This set corresponds to a point in the parameter space where $c g
\to t$ is maximized for non-zero $\delta_{LR}$ (see following
discussion about Table 1 and Table 2) and the results are depicted
in Fig.~\ref{constr6}.
 Since the squark masses are not universal, $\delta_{LR}>1$ can still satisfy all the
constraints and thus is allowed. The reason for this is that all
the constraints actually limit the size of the product $\delta_{LR} M_{Q_2}
M_{U_3}$. For the case discussed here, $M_{Q_2}$ and $M_{U_3}$ are
not large and thus a large $\delta_{LR}$ is allowed.

A common property of the above several figures is that
$B_s-\bar{B}_s$ mixing and $b \to s \gamma$ require a small
$\delta_{LL}$ value. This is a general feature, which accounts for
the significant suppression of the maximal predictions for various
top quark FCNC interactions after considering all the constraints
(see the results in case I($\delta_{LR}=0$) of Table 1) . We also
considered the constraints on $\delta_{RL}$ and $\delta_{RR}$,
which, as we pointed out earlier in this section, do not
significantly affect $b \to s \gamma$ and $B_s-\bar{B}_s$ mixing,
and hence are constrained only by $\delta \rho$ and the Higgs
boson mass. We found that by comparing with the constraints on
$\delta_{LR}$ and $\delta_{LL}$, the constraints on these two
mixing parameters, especially on $\delta_{RR}$, are rather weak.
Let us consider parameters in Eq.~(\ref{susy-para}) as an example.
For $\delta_{LL}=\delta_{LR}=0$, our results indicate that
$\delta_{RR}$ and $\delta_{RL}$ should be less than $0.76$ and
$0.46$, respectively. Such constraints come from $\delta M_W $ and
$\delta \sin^2 \theta_{eff}$, and are sensitive to $X_t$. For $X_t
=0 $,
  the bounds will be relaxed to $0.98$ for
$\delta_{RR}$ and $0.75$ for $\delta_{RL}$.

\subsection{Maximal predictions in MSSM for FCNC top quark processes}

With the constraints discussed above, we perform a scan over
the SUSY parameter space to search for the maximal predictions of the
MSSM on various top quark FCNC processes. We consider two cases: (I) only
$\delta_{LL}\neq 0$ and (II) only $\delta_{LR} \neq 0$. For
each case, we require the parameters to vary in the following
ranges \footnote{There is a typo in Eq.~(15) of Ref.~\cite{cao}:
$\delta_{LL}$ and $\delta_{LR}$ were required to vary between 0 and 2
rather than between 0 and 1. We checked that allowing $\delta$ parameters
to be larger than 2 does not affect the results in Table I, but more samples
are needed to get the results in Table I. In our scan, we required
$\delta_{RL}^d=\delta_{RR}^d=0$ and $\delta_{LL}, \delta_{LR} > 0$.
Relaxing these requirements does not change our results but lowers
the efficiency to search for the maximal predictions in a certain
number of samples. In our scan, we also found that our results are
not sensitive to the values of $M_{D_i}$, $M_{U_1}$, $M_{Q_1}$ and
$A_b$.}
\begin{eqnarray}
&& 2 < \tan\beta < 60, \quad  \quad \quad \quad \quad \quad  0< M_{Q_i,U_i, D_i}<1 {\rm ~TeV}, \nonumber \\
&& 94 {\rm ~GeV} < m_A <1 {\rm ~TeV}, \quad \quad 195 {\rm ~GeV}
<m_{\tilde g} < 1 {\rm ~TeV},
                                                                \nonumber \\
&& 0< \delta_{LL}\ or \ \delta_{LR} < 2, \quad \quad \quad \quad -1 {\rm ~TeV}  < \mu, M_2 < 1 {\rm ~TeV}, \nonumber \\
&& 0< \delta_{LR}^b < 0.1, \quad \quad  \quad \quad \quad \quad \; -2 {\rm ~TeV} < A_{t,b} < 2 {\rm
~TeV} .
\end{eqnarray}
To manifest the effect of the combined constraints on the maximal rates,
 we present two types of predictions: one by only
requiring the squark, chargino and neutralino masses satisfy their
current lower bounds
\begin{eqnarray}
m_{\tilde{u}}> 96  {\rm ~GeV}, \quad m_{\tilde{d}} > 89  {\rm ~GeV}, \quad
m_{\tilde{\chi}^0} > 46  {\rm ~GeV}, \quad m_{\tilde{\chi}^+} > 94  {\rm ~GeV};
\label{mass constraints}
\end{eqnarray}
and the other by imposing all constraints in Eqs.~(\ref{mass
constraints}, \ref{bound1}, \ref{bound2}, \ref{bound3},
\ref{bound4}). With five million samples for each process in
either case, we obtain the maximal predictions and they are given in Table I.  From
the table one can see that the combined constraints can significantly
decrease the MSSM predictions for top-quark FCNC channels at the
LHC, especially for the case I, and among these FCNC processes, $c g
\to t h $ is the most affected one after imposing these constraints. The reason is that,
as we pointed out before, there is a cancellation between the contribution
from the effective interactions $\bar{t} c g$ and $\bar{t} c h$,
and the constraints only allow a region with strong cancellation.

\vspace*{0.2cm}
\noindent {\small Table 1: Maximal predictions for top-quark FCNC
processes
     induced by stop-scharm mixings via gluino-squark loops in the MSSM.
     For the production channels we show the hadronic cross sections at the LHC
     obtained from the given parton-level channels
     and the corresponding charge-conjugate channels.
     For the decays we show the branching ratios. LHC sensitivities listed in the last column are
     for 100fb$^{-1}$ integrated luminosity.}
\vspace*{0.1cm}

\begin{center}
\begin{tabular}{|c|c|c|c|c|c|} \hline
& \multicolumn{2}{c|}{$\delta_{LL}\neq
0$}&\multicolumn{2}{c|}{$\delta_{LR}\neq 0$} & LHC sensitivity \\
\cline{2-5}
        & constraints& constraints& constraints & constraints & at $3 \sigma$ level \\
        &  masses&  all    & masses &   all   &   \\ \hline
 $t \to ch$ & $1.2 \times 10^{-3} $ & $ 2.0 \times
10^{-5}$ & $2.5\times 10^{-2}$ & $ 6.0 \times 10^{-5}$ & $5.8 \times 10^{-5}$ \cite{tch-Aguilar} \\
\hline $t \to c g $ & $5.0 \times 10^{-5}$ & $5.0 \times 10^{-6} $
& $1.3 \times 10^{-4}$ & $3.2 \times 10^{-5}$ & $  - $   \\
\hline $t \to c g g$ & $6.1 \times 10^{-5}$ & $7.1\times 10^{-6}$ & $1.5 \times 10^{-4}$ & $3.5 \times 10^{-5}$ & $-$ \\
\hline $t \to c Z$ & $5.0 \times 10^{-6}$ & $5.7 \times 10^{-7}$
&$1.2 \times 10^{-5}$ &
  $1.8 \times 10^{-6}$                                  &  $3.6 \times 10^{-5}$ \cite{tcz-Han,tcz-Atlas}  \\ \hline
$t \to c \gamma $ & $9.0 \times 10^{-7}$ & $1.5 \times 10^{-7}$ &
$1.3 \times 10^{-6}$ & $5.2 \times 10^{-7}$ & $1.2 \times 10^{-5}$ \cite{tcr-Han,tcr-Beneke} \\
\hline \hline
$cg \to t$ & 1450 fb & 225 fb& 3850 fb & 950 fb         & 800 fb\cite{Hosch} \\
\hline $gg \to t\bar{c}$ & 1400 fb & 240 fb & 2650 fb & 700 fb
 & 1500 fb \cite{pptc-Han,pptc-background}
\\ \hline $ c g \to tg$ & 800 fb & 85 fb & 1750 fb & 520 fb
& 1500 fb \cite{pptc-Han,pptc-background}\\ \hline $cg \to t
\gamma$ & 4 fb & 0.4 fb& 8 fb & 1.8 fb & 5 fb \cite{zt-Aguilar}
\\ \hline $cg \to tZ$ & 11 fb & 1.5 fb & 17 fb & 5.7 fb & 35 fb \cite{zt-Aguilar}  \\
\hline $c g \to th$ & 550 fb & 18 fb & 12000 fb & 24 fb & 200 fb \cite{tch-Aguilar}  \\
\hline
\end{tabular}
\end{center}

\vspace*{0.2cm}
\begin{center}
\noindent {\small Table 2:  SUSY parameters leading to the maximal
predictions for $\delta_{LR}\neq 0$ in Table 1.} \vspace*{0.2cm}

\vspace*{0.1cm}
\begin{tabular}{|c|c|c|c|c|c|c|} \hline
process & $M_{Q_2} (GeV) $  &$M_{Q_3} (GeV) $ & $M_{U_3} (GeV) $ &
$X_t (GeV) $ & $m_{\tilde{g}} (GeV) $ &  $\delta_{LR}$ \\ \hline
\hline $t \to ch$ & 1000 & 900 & 225 & 500 & 195 & \ \ 1.1\ \  \\
\hline $t \to c g $ & 310 & 985 & 90 & 100  & 195 & \ \ 1.72\ \  \\
\hline $t \to c g g$ & 310 & 980 & 90 & 35  & 195 & \ \ 1.75\ \  \\
\hline $t \to c Z$ & 500 & 900 & 165 & 600 & 198 & \ \ 1.1 \ \
\\ \hline
$t \to c \gamma $ & 290 & 1000 & 90 & 20 & 196 & \ \ 1.72\ \  \\
\hline \hline
$cg \to t$ & 400 & 990 & 120 & 33 & 196 & 1.5 \\
\hline $gg \to t\bar{c}$ & 310 & 990 & 85 & 80 & 196 & \ \ 1.8 \ \
\\ \hline $ c g \to tg$ & 490 & 900 & 125 & 0 & 197 & \ \ 1.45 \ \  \\
 \hline $cg \to t \gamma$ & 280 & 1000 & 85 & 25 & 197 & \ \ 1.78
 \ \
 \\ \hline $cg \to tZ$ & 370  & 920 & 80 & 115 & 196 & \ \ 1.86 \ \ \\
\hline $c g \to th$ & 280 & 1000 & 85 & 23 & 197 & \ \ 1.77\ \ \\
\hline
\end{tabular}
\end{center}

As one can see from Table 1, predictions for the case with
non-zero $\delta_{LR}$ are larger than for the one with non-zero
$\delta_{LL}$. In Table 2 we list the SUSY parameters leading to the
maximal predictions in Table 1 for the case II ($\delta_{LR}\neq
0$).  It is seen from Table 2 that the `favorable' parameters for
the maximal rates are process dependent. Since these parameters
could be first tested by seeking top quark FCNC signals at LHC, in
Table 3 we present the predictions for all processes with two sets
of parameters, called `Point 1' and `Point 2', where $t\to c h$ and
$c g \to t $ are required to be maximized, respectively. It is seen
that `Point 1'  favors $cg\to t$ as the production channel but
`Point 2' favors $t\to cgg$ among the decay modes.

Note that in Table 1 we only showed the cases of $\delta_{LL}\neq 0$
and $\delta_{LR} \neq 0$. For $\delta_{RL}\neq 0$, we found that the
maximal rate of $t \to cg$ is $1.3 \times 10^{-4}$ if only the
squark mass constraints are included and $6 \times 10^{-5}$ with all
the constraints. For $\delta_{RR} \neq 0$, the maximal rate of $t
\to cg$ is $5.0 \times 10^{-5}$ with only the squark mass
constraints and $4.85 \times 10^{-5}$ with all the constraints.
These results can be easily understood since, as discussed in
\cite{cao} and in this section, the constraints on $\delta_{RL}$ and
$\delta_{RR}$ are weaker than those for $\delta_{LR}$ and
$\delta_{RR}$, respectively.

\vspace*{0.2cm} \noindent {\small Table 3: SUSY predictions for the
rates of top quark FCNC processes at two points of parameter space:
`Point 1' maximizes $t \to c h$ and `Point 2' maximizes $c g \to t $
channel. }

\vspace*{0.1cm}
\begin{center}
\begin{tabular}{|c|c|c|} \hline
process & Point 1 & Point 2  \\ \hline $t \to c g$ & $2.0 \times
10^{-5} $ & $ 3.1 \times 10^{-5}$ \\ \hline $t \to c g g $ & $2.9
\times 10^{-5} $ & $ 3.5 \times 10^{-5}$ \\ \hline $t \to c Z$ &
$1.4 \times 10^{-6} $ & $ 1.2 \times 10^{-6}$  \\ \hline $t \to c
\gamma $ & $2.6 \times 10^{-7}$& $ 5.2 \times 10^{-7}$ \\ \hline
$t \to c h$ & $6.0 \times 10^{-5} $ & $ 7.0 \times 10^{-9}$ \\
\hline $c g \to t$ & 660 fb & 950 fb \\ \hline $g  g \to t \bar{c}$
& 450 fb & 690 fb \\ \hline $c g \to t g $ & 280 fb & 415 fb \\
\hline $c g \to t Z$ & 2.3 fb & 4.9 fb \\ \hline $c g \to t \gamma $
& 1.2 fb & 1.8 fb \\ \hline $c g \to t h $ & 24 fb & 8.5 fb \\
\hline
\end{tabular}
\end{center}

\subsection{Top FCNC observability at the LHC}
Now we consider the observability of these top quark FCNC
interactions at the LHC. This issue has been intensively investigated
in the effective Lagrangian  approach  for $t \to c
h$ \cite{tch-Aguilar}, $t \to c Z$ \cite{tcz-Han,tcz-Atlas}, $t
\to c \gamma$ \cite{tcr-Han,tcr-Beneke}, $p p \to t +X$
\cite{Hosch}, $ p p \to t \bar{c}+X$
\cite{pptc-Han,pptc-background}, $p p \to t g +X
$\cite{pptc-Han,pptc-background}, $ p p \to t
Z+X$\cite{zt-Aguilar}, $p p \to t \gamma +X $\cite{zt-Aguilar} and
$p p \to t h +X$\cite{tch-Aguilar}. At the
LHC, most of the top quarks will be pair-produced. One of the tops is
assumed to subsequently decay as $t\to bW$ (normal mode) while
the other top goes to
one of the  above channels via FCNC interactions (exotic mode).

Due to the
large QCD backgrounds at the LHC, the search for these processes
must be performed in the decay channels $W \to \ell
\bar{\nu}_\ell$ ($\ell=e,\mu$) for $W$ boson, $Z\to \ell^+ \ell^-$
for $Z$ boson and $ h \to b \bar{b}$ for Higgs boson. For any of
these reactions, top quark reconstruction is required to extract the
signal from its background. In Table 4, we list the signals and
the main backgrounds. The detailed Monte Carlo
simulations for the signals and backgrounds can be found in the
corresponding literature listed in the last column of Table 1,
where the LHC sensitivity for these processes is quoted.
Although these sensitivities are based on the effective Lagrangian
approach and may be not perfectly applicable to the MSSM, we can
take them as a rough criteria to estimate the observability of
these channels. Comparing these sensitivities with the SUSY
predictions, one can see that only the maximal predictions for $c
g \to t $ and $ t \to c h $ are slightly larger than the
corresponding LHC sensitivities. This implies that the study of these
processes may provide the first insight about top quark FCNC. From
these sensitivities one can also see that, although the $ g g \to t
\bar{c}$ channel is now the most extensively studied among the FCNC
production channels \cite{pptc-mssm}, its observation needs a
higher luminosity. This is because $ g g \to t \bar{c}$ has large
irreducible backgrounds from single top productions in the SM
\cite{pptc-background}. Note that in Table 1, we did not list the
sensitivity of the LHC to $ t \to c g$, which has not been
investigated because of a general belief that this decay is not well
suited for detecting $\bar{t} c g$ interaction
\cite{Aguilar-Saavedra:2004wm}. In fact, this decay was once
investigated for the Tevatron \cite{tcg-Han}, and it was found
that due to the large background, namely $W$ boson plus three
jets, only a branching ratio as large as $5 \times 10^{-3}$ may be
accessible with 10 fb$^{-1}$ integrated luminosity. This rate is
about one order larger than that for $ t \to c \gamma$ at the
Tevatron with the same luminosity \cite{tcr-Han}. For the decay $t
\to c g g$, there is an additional jet in its signal and its
observability needs to be studied by a detailed Monte Carlo
simulation.

\vspace*{0.3cm} \noindent {\small Table 4: Experimental signature
and main background for FCNC top rare decays and productions at
the LHC. The top quarks are assumed to decay $t \to W^+ b \to
\ell^+ \nu_\ell b$, and $Z$ and $h$ bosons decay in the channel $Z
\to \ell^+ \ell^-$ and $h \to b \bar b$, respectively.}
\vspace*{0.2cm}

\begin{center}
\begin{tabular}{lccclcc}
\hline Process & Signal & Background & & ~~~~Process &  Signal &
Background
\\ \hline
$t \bar t$, $t \to c g$ & $j j \ell \nu b$ & $Wjjj$ & & ~~~~$c g \to
t$ & $\ell \nu b$ & $Wj$  \\ \hline $t \bar t$, $t \to c g g$ & $j j
j \ell \nu b$ & $Wjjj j$ && ~~~~$g g \to t \bar{c}$ & $\ell \nu b j$
& $t j$  \\ \hline  & & & & ~~~~$c g \to t g$ & $\ell \nu b j $ & $
t j $
\\  \hline $t \bar t$, $t \to c Z$ &$\;\;$ $\ell^+ \ell^- j \ell \nu
b$ & $ZWjj$ && ~~~~$c g \to t Z$ &$\;\;$ $\ell^+ \ell^- \ell \nu b$
& $ZWj$  \\ \hline $t \bar t$, $t \to c \gamma$ & $\gamma j \ell \nu
b$ & $\gamma Wjj$ & & ~~~~$c g \to t \gamma $ & $\gamma \ell \nu b$
& $\gamma Wj$  \\ \hline $t \bar t$, $t \to c h$ & $b \bar b j \ell
\nu b$ & $Wb \bar b jj$ & &
~~~~$c g \to t h$ & $b \bar b \ell \nu b$ & $t \bar t$  \\
\hline
\end{tabular}
\end{center}

\section{Conclusions}
In this paper, we  investigated systematically the SUSY-induced top
quark FCNC processes at the LHC, which includes various decay modes
and production channels. We performed a comparative study for all
the decay modes and for all the production channels so that one can
see clearly which decay mode or production channel can have a
relatively large rate. The dependence of these channels
on the relevant SUSY parameters is investigated in detail and its
properties  are analyzed. We note that such a global study of the
top quark FCNC processes has been done only in a model independent
way \cite{Aguilar-Saavedra:2004wm}. We also analyzed the
characteristics of the `favorable region' in SUSY parameter space
where the FCNC processes are maximized. After getting an
understanding of these processes, we examined the effects of all the
direct and indirect experimental constraints on the scharm-stop
flavor mixings and scanned the parameter space to find their maximal
rates with these constraints imposed. We found that $ c g \to t$ and
$ t \to c h$ are the most likely channels to be observable at the
LHC if the MSSM is the correct scenario beyond the SM.

\section*{Acknowledgment}

This work is supported in part by a fellowship from the Lady Davis
Foundation at the Technion, by the Israel Science Foundation (ISF), by NSERC
of Canada under Grant No. SAP01105354,
the National Natural Science
Foundation of China under Grant No. 10475107 and 10505007,  the
Grant-in-Aid for Scientific Research (No. 14046201) from the Japan
Ministry of Education, Culture, Sports, Science and Technology,
and by the IISN and the Belgian science policy office (IAP V/27).

\appendix

\section{Expressions for the Loop results}

If significant flavor mixings exist only between left-handed
scharm with stops, the squark states $\tilde{c}_L$, $ \tilde{t}_L$
and $\tilde{t}_R$ will mix together to induce various top quark
FCNCs. In this case, other squark states only serve as spectators
to the processes considered in this paper. So in the actual
calculation, we only need to consider the squark mass matrix for
$(\tilde{t}_L, \tilde{t}_R, \tilde{c}_L)$, which is given by
\begin{eqnarray}
\left ( \begin{array}{ccc}
    M_{Q_3}^2 + m_t^2 + (\frac{1}{2} - \frac{2}{3} s_W^2 ) m_Z^2 \cos 2 \beta  &  m_t ( A_t -\mu \cot \beta)&  \delta_{LL} M_{Q_2} M_{Q_3} \\
     m_t ( A_t -\mu \cot \beta)  &  M_{U_3}^2 + m_t^2 + \frac{2}{3} s_W^2 \cos 2 \beta  &  \delta_{LR} M_{Q_2} M_{U_3} \\
    \delta_{LL} M_{Q_2} M_{Q_3}    &  \delta_{LR} M_{Q_2} M_{U_3} & M_{Q_3}^2 + (\frac{1}{2} - \frac{2}{3} s_W^2 ) m_Z^2 \cos 2 \beta  \end{array}  \right
    ).
\end{eqnarray}
This mass matrix can be diagonalized by a unitary matrix $V$, and
it enters the squark interactions as in Eq.~(\ref{interaction}).

In this Appendix, we list the expressions for $\Sigma$ and
$\Gamma^{\bar{t}c g}_\mu $ in Eq.~(\ref{eff}) which are needed to
get the effective $\bar{t} c g $ vertex. We also list the
expressions for $\Gamma^{\bar{t}c \gamma} $, $\Gamma^{\bar{t} c Z}
$ and $\Gamma^{\bar{t}c h} $ to calculate other effective
vertices. Before presenting these expressions, we define the
following abbreviations:
\begin{eqnarray}
R^a & = &  \sum_{\lambda=1}^3 V_{1 \lambda} V_{\lambda 3}^\dag
B(p, m_{\tilde{g}},  m_{\lambda}), \quad R^b = R^a |_{V_{1
\lambda} \to V_{2 \lambda}},
\end{eqnarray}
\begin{eqnarray}
R^c & = &  \sum_{\lambda=1}^3 V_{1 \lambda} V_{\lambda 3}^\dag
C(-p_c, p_c - p_t,  m_{\tilde{g}},  m_{\lambda},  m_{\lambda} ),
\quad R^d = R^c |_{V_{1 \lambda} \to V_{2 \lambda}},
\end{eqnarray}
\begin{eqnarray}
R^e & = &  \sum_{\lambda=1}^3 V_{1 \lambda} V_{\lambda 3}^\dag
C(-p_c, p_c - p_t,  m_{\lambda},  m_{\tilde{g}},  m_{\tilde{g}}),
\quad R^f = R^e |_{V_{1 \lambda} \to V_{2 \lambda}},
\end{eqnarray}
\begin{eqnarray}
R^g & = &  \sum_{\lambda=1}^3 V_{1 \lambda} V_{\lambda 3}^\dag
C(-p_b, p_b - p_t,  m_{\tilde{g}},  m_{\tilde{b}_L},  m_{\lambda}
),
\end{eqnarray}
\begin{eqnarray}
R^h & = &  \sum_{\rho, \lambda=1}^3 V_{1 \lambda} F^Z_{\lambda
\rho} V_{\rho 3}^\dag C(-p_c, p_c - p_t,  m_{\tilde{g}},
m_{\rho},  m_{\lambda} ), \quad R^i = R^h |_{V_{1 \lambda} \to
V_{2 \lambda}},
\end{eqnarray}
\begin{eqnarray}
R^j & = &  \sum_{\rho, \lambda=1}^3 V_{1 \lambda} F^h_{\lambda
\rho} V_{\rho 3}^\dag C(-p_c, p_c - p_t,  m_{\tilde{g}},
m_{\rho},  m_{\lambda} ), \quad R^k = R^j |_{V_{1 \lambda} \to
V_{2 \lambda}},
\end{eqnarray}
\begin{eqnarray}
R^l & = &  \sum_{\lambda=1}^3 V_{1 \lambda} V_{\lambda 3}^\dag
D(-p_t, p_2, p_1, m_{\lambda}, m_{\tilde{g}}, m_{\tilde{g}},
m_{\tilde{g}} ), \quad R^m = R^l |_{V_{1 \lambda} \to V_{2
\lambda}},
\end{eqnarray}
\begin{eqnarray}
R^n & = &  \sum_{\lambda=1}^3 V_{1 \lambda} V_{\lambda 3}^\dag D(
p_1, p_2, -p_t, m_{\lambda}, m_{\lambda}, m_{\lambda},
m_{\tilde{g}} ),
 \quad R^o = R^n |_{V_{1 \lambda} \to V_{2 \lambda}},
\end{eqnarray}
\begin{eqnarray}
R^p & = &  \sum_{\lambda=1}^3 V_{1 \lambda} V_{\lambda 3}^\dag D(
-p_t, p_2, - p_c, m_{\lambda}, m_{\tilde{g}}, m_{\tilde{g}},
m_{\lambda} ), \quad R^q = R^p |_{V_{1 \lambda} \to V_{2
\lambda}},
\end{eqnarray}
where $\rho$ and $\lambda$ are squark indices in mass eigenstate,
$p_i$ is particle momentum, B, C and D are loop
functions\cite{Hooft}, and $F^Z_{\lambda \rho}$ and $F^h_{\lambda
\rho}$ are interaction coefficients for $\tilde{q}_\lambda^*
\tilde{q}_\rho Z $ and $\tilde{q}_\lambda^* \tilde{q}_\rho h $
interactions respectively, which are given by
\begin{eqnarray}
F^Z_{\lambda \rho} &= & (1 -\frac{4}{3} s_W^2 ) \delta_{\rho
    \lambda} - V_{\lambda 2}^\dag V_{2 \rho}   \\
F^h_{\lambda \rho} & = & -\frac{g}{m_W \sin \beta } \left ( m_t^2
   \cos \alpha ( V_{\lambda 1}^\dag V_{1 \rho} + V_{\lambda 2}^\dag
V_{2 \rho} ) + \frac{m_t ( A_t \cos \alpha + \mu \sin \alpha )}{2}
(V_{\lambda 1}^\dag V_{2 \rho} +
V_{\lambda 2}^\dag V_{1 \rho}) \right. \nonumber\\
& & \left .  + \frac{\delta_{LR} M_{Q_2} M_{U_3} \cos \alpha}{2}
(V_{\lambda 3}^\dag V_{2 \rho} + V_{\lambda 2}^\dag V_{3 \rho})
\right ).
\end{eqnarray}
Then after factoring out the common factor $\alpha_s/4 \pi$, we
obtain the expressions for $\Sigma$ and $\Gamma $s:
\begin{eqnarray}
\Sigma (p) &=& - 2 C_F \left ( \gamma^\mu R^a_\mu + m_{\tilde{g}}
                 R^b_0 \right ) P_L,   \\
\Gamma^{\bar{t} c g}_\mu &=& \frac{1}{3} g_s T^a \left ( 2
  \gamma^\nu R^c_{\mu \nu} - 2 m_{\tilde{g}} R^d_\mu - (p_t +
p_c)_\mu (\gamma^\nu R^c_\nu - m_{\tilde{g}} R^d_0 ) \right ) P_L
 + 3 g_s T^a \left ( \p_slash_t \gamma_\mu \p_slash_c R^e_0 -
\p_slash_t \gamma_\mu \gamma^\nu R^e_\nu  \right .  \nonumber\\
&& \left . - \gamma^\nu \gamma_\mu \p_slash_c R^e_\nu + \gamma^\nu
\gamma_\mu \gamma^\lambda R^e_{\nu \lambda} + m_{\tilde{g}}^2
\gamma_\mu R^e_0 - m_{\tilde{g}} \gamma_\mu \p_slash_c R^f_0 + 2
m_{\tilde{g}} R^f_\mu - m_{\tilde{g}} \p_slash_t \gamma_\mu R^f_0
\right ) P_L,    \\
\Gamma^{\bar{b} c W}_\mu &=& \frac{g}{\sqrt{2}} C_F \gamma^\nu P_L
(p_{b \mu} R^g_\nu - R^g_{\mu \nu} ),  \\
\Gamma^{\bar{t} c Z}_\mu &=& \frac{g}{c_W} C_F \left ( (p_t + p_c
)_\mu ( \gamma^\nu R^h_\nu - m_{\tilde{g}} R^i_0 ) - 2 (
\gamma^\nu R^h_{\mu \nu} - m_{\tilde{g}} R^i_\mu ) \right ) P_L, \\
\Gamma^{\bar{t} c \gamma}_\mu &=& - 2 e Q_u C_F \left ( 2
\gamma^\nu R^c_{\mu \nu} - 2 m_{\tilde{g}} R^d_\mu - (p_t +
p_c)_\mu (\gamma^\nu R^c_\nu - m_{\tilde{g}} R^d_0 ) \right ) P_L,\\
\Gamma^{\bar{t} c h} &=&  -2 C_F \left ( \gamma^\mu R^j_\mu -
m_{\tilde{g}} R^k_0 \right ) P_L,
\end{eqnarray}
where $C_F$ is the quadratic Casimir operator of the fundemantal
representation of $SU(3)_C$.

For the box diagrams in Fig.~\ref{SQCD diagram}, their results are
given by
\begin{eqnarray}
M_d &=& -2 \alpha_s^2 T^e T^a f^{d e b} f^{c b a}
\varepsilon^d_\lambda (p_2) \varepsilon^c_\rho (p_1) \bar{u}_t
\left ( - m_{\tilde{g}} m_t \gamma^\lambda
(\p_slash_t - \p_slash_2 ) \gamma^\rho R^m_0  \right.\nonumber \\
&& +  m_{\tilde{g}} m_t \gamma^\lambda \gamma^\nu \gamma^\rho
R^m_\nu + m_{\tilde{g}} \gamma^\nu \gamma^\lambda (\p_slash_t -
\p_slash_2 ) \gamma^\rho R^m_\nu - m_{\tilde{g}} \gamma^\mu
\gamma^\lambda \gamma^\nu
\gamma^\rho R^m_{\mu \nu}  \nonumber \\
&&  \quad \   - m_t \gamma^\lambda (\p_slash_t - \p_slash_2 )
\gamma^\rho \gamma^\nu R^l_\nu + m_t \gamma^\lambda \gamma^\mu
\gamma^\rho \gamma^\nu R^l_{\mu \nu} + \gamma^\mu \gamma^\lambda
(\p_slash_t - \p_slash_2 )  \gamma^\rho \gamma^\nu R^l_{\mu \nu} -
\gamma^\mu \gamma^\lambda \gamma^\nu \gamma^\rho \gamma^\sigma
R^l_{\mu \nu \sigma} \nonumber \\
 && \left . \quad \  + m_{\tilde{g}}^2 m_t \gamma^\lambda \gamma^\rho R^l_0 -
 m_{\tilde{g}}^2 \gamma^\nu \gamma^\lambda \gamma^\rho R^l_\nu +
 m_{\tilde{g}} m_t \gamma^\lambda \gamma^\rho \gamma^\nu R^m_\nu
 - m_{\tilde{g}} \gamma^\mu \gamma^\lambda \gamma^\rho \gamma^\nu R^m_{\mu \nu} \right ) P_L
 v_c,  \\
M_e &=& - 8 \alpha_s^2 T^a T^d T^c T^a \varepsilon^d_\lambda (p_2)
\varepsilon^c_\rho (p_1) \bar{u}_t \left (  \gamma^\nu R^n_{\nu
\rho \lambda} + \gamma^\nu p_{1 \lambda} R^n_{\nu \rho} +
m_{\tilde{g}} R^o_{\rho \lambda} + m_{\tilde{g}} p_{1 \lambda}
R^o_\rho \right ) P_L v_c,  \\
M_f &=& - 4 \alpha_s^2 T^a T^c T^b f^{d a b} \varepsilon^d_\lambda
(p_2) \varepsilon^{c}_\rho (p_1)  \bar{u}_t
\left ( m_t \gamma^\lambda (\p_slash_t - \p_slash_2 )R^p_\rho  \right.\nonumber \\
&& - m_t \gamma^\lambda \gamma^\nu R^p_{\nu \rho} - \gamma^\nu
\gamma^\lambda  (\p_slash_t - \p_slash_2 ) R^p_{\nu \rho} +
\gamma^\mu \gamma^\lambda \gamma^\nu R^p_{\mu \nu \rho} -
m_{\tilde{g}} m_t \gamma^\lambda R^q_\rho \nonumber \\
 && \left . \quad \  + m_{\tilde{g}} \gamma^\nu \gamma^\lambda
 R^q_{\rho \nu} - m_{\tilde{g}} \gamma^\lambda (\p_slash_t - \p_slash_2
 ) R^q_\rho + m_{\tilde{g}} \gamma^\lambda \gamma^\nu R^q_{\rho
 \nu} + m_{\tilde{g}}^2 \gamma^\lambda R^p_\rho \right ) P_L v_c.
\end{eqnarray}

\begingroup\raggedright
\endgroup

\end{document}